\documentclass[letterpaper, 11pt]{article}

\usepackage[nocompress]{cite}
\usepackage{jheppub}

\usepackage{graphicx}
\usepackage{epstopdf}
\usepackage{amsmath, amssymb}

\usepackage{bm}
\usepackage{bbold}
\usepackage{caption}
\usepackage{subcaption}

\usepackage{multirow} 
\usepackage{longtable} 

\usepackage{ytableau}

\newcommand{\be}{\begin{eqnarray}}
\newcommand{\ee}{\end{eqnarray}}

\newcommand{\bn}{\begin{enumerate}}
\newcommand{\en}{\end{enumerate}}

\def\IC{\mathbb{C}}

\def\IZ{\mathbb{Z}}

\def\CM{{\cal M}}
\def\CN{{\cal N}}

\def\CR{{\cal R}}
\def\CS{{\cal S}}
\def\CT{{\cal T}}

\def\a{\alpha}

\def\s{\sigma}

\def\G{\Gamma}

\def\half{\frac{1}{2}}

\def\vev#1{\langle #1 \rangle}

\def\tr{{\rm Tr}}

\newcommand{\bea}{\begin{eqnarray}}
\newcommand{\eea}{\end{eqnarray}}

\def\IC{\mathbb{C}}

\def\IZ{\mathbb{Z}}

\def\CM{{\cal M}}
\def\CN{{\cal N}}

\def\CR{{\cal R}}
\def\CS{{\cal S}}
\def\CT{{\cal T}}

\def\a{\alpha}

\def\s{\sigma}

\def\G{\Gamma}

\def\half{\frac{1}{2}}

\def\vev#1{\langle #1 \rangle}

\def\tr{{\rm Tr}}

\title{$\CN=1$ Lagrangians for generalized Argyres-Douglas theories}

\author[a]{Prarit Agarwal,}
\author[b]{Antonio Sciarappa,}
\author[b]{and Jaewon Song}
\affiliation[a]{Department of Physics and Astronomy \& Center for Theoretical Physics\\ Seoul National University, Seoul 08826, Korea}
\affiliation[b]{School of Physics, Korea Institute for Advanced Study\\ 85 Hoegiro, Dongdaemun-gu, Seoul 02455, Korea }

\emailAdd{agarwalprarit@gmail.com}
\emailAdd{asciara@kias.re.kr}
\emailAdd{jsong@kias.re.kr}

\abstract
{
We find $\CN=1$ Lagrangian gauge theories that flow to generalized Argyres-Douglas theories with $\CN=2$ supersymmetry. We find that certain SU quiver gauge theories flow to generalized Argyres-Douglas theories of type $(A_{k-1}, A_{mk-1})$ and $(I_{m, k m}, S)$. We also find quiver gauge theories of SO/Sp gauge groups flowing to the $(A_{2m-1}, D_{2mk+1})$, $(A_{2m}, D_{2m(k-1)+k})$ and $D_{m(2k+2)}^{m(2k+2)}[m]$ theories. 
}

\preprint{SNUTP17-003, KIAS-P17053}

\begin{document}
\maketitle

\section{Introduction}

Argyres-Douglas (AD) theories are strongly-coupled $\CN=2$ superconformal field theories that have (at least one) fractional dimensional Coulomb branch operators. Originally, AD theory was discovered as a low-energy limit of the effective field theory describing the special loci of the Coulomb branch of an $\CN=2$ gauge theory \cite{Argyres:1995jj,Argyres:1995xn}. At this special  loci, we have massless mutually non-local electromagnetically charged particles which are described by an interacting conformal field theory. 
There has been many generalizations of the original construction \cite{Eguchi:1996ds,Eguchi:1996vu, Cecotti:2010fi,Xie:2012hs,Cecotti:2012jx,Cecotti:2013lda,Wang:2015mra,Xie:2016uqq,Xie:2017vaf}. 

Since AD theories have non-integer Coulomb branch operators, the conformal points cannot be described by $\CN=2$ Lagrangian gauge theories. Moreover, writing a Lagrangian for a quantum field theory of both electrically and magnetically charged particles has been a long-standing problem, which has been achieved only by sacrificing manifest Lorentz symmetry or a second-quantized picture. 
This lack of a Lagrangian description posed some challenges in understanding conformal phase of AD theories.  This difficulty has been partially overcome by considering $\CN=1$ Lagrangian gauge theories that flow to Argyres-Douglas theories \cite{Maruyoshi:2016tqk, Maruyoshi:2016aim, Agarwal:2016pjo}. 

Such $\CN=1$ gauge theories were constructed via certain $\CN=1$ preserving deformations of an $\CN=2$ superconformal field theory $\CT_{UV}$ labelled by a nilpotent element of the flavor symmetry algebra $\mathfrak{g}$:
\begin{align}
 \CT_{UV} \leadsto \CT_{IR}[\CT_{UV}, \rho] \ , 
\end{align}
where $\rho$ is an embedding of $\mathfrak{su}(2)$ into $\mathfrak{g}$ labelling the nilpotent element of $\mathfrak{g}$. The commutant of $\rho$ determines the remaining flavor symmetry. 
We deform the UV theory by introducing a chiral multiplet $M$ that transforms as an adjoint of $\mathfrak{g}$ which is coupled via a superpotential term
\begin{align}
 W = \tr M \mu \ .
\end{align} 
Here $\mu$ is the moment map operator for the flavor symmetry $\mathfrak{g}$ which also transforms as the adjoint. Then we give a nilpotent vacuum expectation value $\vev{M} = \rho(\s^+)$ to the chiral multiplet, which triggers an RG flow to an $\CN=1$ theory. 
This type of deformation was first considered in \cite{Gadde:2013fma} and was also later studied in \cite{Agarwal:2013uga,Agarwal:2014rua,Giacomelli:2014rna, Agarwal:2015vla,Fazzi:2016eec,Nardoni:2016ffl}. 

One crucial feature of this deformation is that it preserves a $U(1)$ symmetry that can be mixed with the $R$-symmetry. The superconformal $R$-symmetry in the IR is determined via $a$-maximization \cite{Intriligator:2003jj}. Generically, there exists an accidental symmetry caused by some operators hitting the unitarity bound along the RG flow. These unitarity-violating operators become free and get decoupled. We need to subtract this piece and do the $a$-maximization again to correctly obtain the superconformal $R$ \cite{Kutasov:2003iy}. We need to repeat this until all the operators have dimension above the unitarity bound. 

Now, if we choose $\CT_{UV}$ to be the conformal SQCD such as $SU(N)$ gauge theory with $2N$ fundamental hypermultiplets and $\rho$ to be principal embedding that breaks the $SU(2N)$ symmetry completely (but baryonic $U(1)$ and axial $U(1)$ symmetry are left unbroken), the $\CT_{IR}$ is given by $(A_1, A_{2N-1})$ Argyres-Douglas theory, upon removing operators that become free along the RG flow.  Since the deformation will give a nilpotent mass to the fundamental hypermultiplets, we can integrate them out. This produces an $\CN=1$ gauge theory that flows to the $\CN=2$ AD theory, which provides a new handle to investigate this strongly-coupled SCFT.  
It was also found that if we dimensionally reduce the setup to 3d, similar SUSY enhancing RG flow can be obtained once we impose additional constraint to remove the operators that would be removed in 4d \cite{Benvenuti:2017kud,Benvenuti:2017lle}. 

The $\CN=1$ Lagrangian gauge theories that flow to AD theories have many interesting applications. First of all, one can use localization to compute various supersymmetric partition functions.  The full superconformal index of the $\CN=2$ fixed point was computed in \cite{Maruyoshi:2016tqk,Maruyoshi:2016aim,Agarwal:2016pjo}. 
This result has been checked against a number of independent computations in specialized limits. Realization of the AD theories in terms of M5-branes makes it possible to compute the Schur and Macdonald indices using the TQFT living on a Riemann surface with an irregular puncture \cite{Buican:2015ina, Buican:2015tda,Song:2015wta,Buican:2017uka, Song:2017oew}. Also, it is possible to use the 3d mirror description to compute the Hall-Littlewood index \cite{DelZotto:2014kka}. 
There is also an interesting connection between the Schur index and the spectrum of massive BPS particles in the Coulomb branch \cite{Cecotti:2015lab,Cordova:2015nma,Cordova:2016uwk, Cordova:2017ohl}. 
The correspondence between 4d $\CN=2$ SCFT and the 2d chiral algebra \cite{Beem:2013sza} gives another way of computing the Schur/Macdonald index \cite{Cordova:2015nma, Xie:2016evu,Song:2016yfd,Creutzig:2017qyf,Song:2017oew} of the Argyres-Douglas theories. The Lagrangian description would be helpful to understand the surface defect of the AD theories \cite{Cordova:2017mhb} as well. Finally, the Lagrangian description for the AD theory has been used to obtain the Coulomb branch Lens space index \cite{Fredrickson:2017yka} which computes a wild Hitchin character. It was also used to topologically twist the AD theory \cite{Gukov:2017zao}. 

In this paper, we extend the construction of \cite{Maruyoshi:2016tqk, Maruyoshi:2016aim, Agarwal:2016pjo} for the $(A_1, A_n)=(I_{2, n+1})$ and $(A_1, D_n)=(I_{2, n-2}, S)$ to more generalized AD theories that are called $(A_{k-1}, A_{mk-1}) = (I_{k, mk})$ and $(I_{m, mk}, S)$ theories (or $A_{mk-1}^{mk}[m]$ and $A_{mk}^{mk}[m]$ in the notation of \cite{Wang:2015mra,Xie:2016evu}). We find that conformal quiver gauge theories of certain type flow to AD theories upon $\CN=1$ deformations with $\rho$ being the principal embedding. 
The relevant quivers are as follows:
\begin{align}
\label{eq:SUquiver}
\begin{array}{ccl}
 (N)-(2N)-(3N)-\ldots-(mN-N)-\boxed{mN} & \leadsto & (A_{m-1}, A_{Nm-1}) \\
 \boxed{1}-(k+1)-(2k+1)-\ldots-(mk-k+1)-\boxed{mk+1} & \leadsto & (I_{m, mk}, S) 
\end{array}
\end{align}
Here nodes in the parenthesis denote $SU$ gauge groups, while the nodes in the box denote the flavor $U$ group. We find that once we perform the nilpotent deformation for the flavor $SU(mN)$ or $SU(mk+1)$ on the right-hand side using the principal embedding, we obtain the Argyres-Douglas theory in the IR upon removing decoupled operators. Here we do not touch the $U(1)$ factors of the flavor group. In the end we have $U(1)^{m-1}$ and $U(1)^m$ flavor symmetry for the first and second quiver theory upon principal deformation. 

We also find that the following quiver gauge theories made out of $SO$ and $Sp$ gauge group\footnote{Our notation for the symplectic group is chosen so that $Sp(1)=SU(2)$.}  flow to Argyres-Douglas type theories (with $N$ even for \eqref{eq:SOSP1}):\footnote{Interestingly, in all cases in which the IR Argyres-Douglas theory we obtain after the $\mathcal{N} = 1$ deformation is of $(G,G')$-type with $G,G'$ simply-laced Lie algebras, we find that 
the Dynkin diagram of $G$ is realized by the gauge nodes of the UV quiver theory (before the deformation), while $G'$ coincides with the Lie algebra of the non-Abelian flavor symmetry group of the UV quiver theory.
}$^{,}$\footnote{While this paper was undergoing review by JHEP, \cite{Giacomelli:2017ckh} appeared on arXiv where the author independently identified the AD theory corresponding to the IR SCFT obtained from the aforementioned $\mathcal{N}=1$ deformation of the quiver shown on the LHS of \eqref{eq:SPSO1}.}
\begin{align}
\label{eq:2ndSPSO}
\begin{split}
\begin{array}{ccl}
 \boxed{SO(2)}-Sp(N)-SO(4N+2)-Sp(3N)-\\
 ~~~~~~~~~~~~~~~~~~~~~~~ \ldots-Sp(2mN-N)-\boxed{SO(4mN+2)} 
 &  \leadsto & (A_{2m-1}, D_{2Nm+1})
\end{array}
\end{split}
\end{align}

\begin{align}
\label{eq:SOSP1}
\begin{split}
\begin{array}{ccl}
  SO(N)-Sp(N-2)-SO(3N-4)-Sp(2N-4)- \\
 ~~~~~~~~~~~~~~~~ 
 \ldots-Sp(m(N-2))-\boxed{SO(2m(N-2)+N)}  
 &  \leadsto & (A_{2m}, D_{m(N-2)+\frac{N}{2}})
\end{array}
\end{split}
\end{align}

\begin{align}
\label{eq:SPSO1}
\begin{split}
\begin{array}{ccl}
  Sp(N)-SO(4N+4)-Sp(3N+2)-SO(8N+8)- ~~~~~~~~~~~~\\
~~~~~~~~~~~~~~   \ldots-Sp((m-1)(2N+2)+N)-\boxed{SO(4m(N+1))}  
 &  \leadsto & D_{m(2N+2)}^{m(2N+2)}[m]
 \end{array}
\end{split}
\end{align}
Here $(A_{2m-1}, D_{2Nm+1}) = D_{2Nm+1}^{4Nm}[2m]$ and $(A_{2m}, D_{m(N-2)+\frac{N}{2}}) = D_{(2m+1)(\frac{N}{2}-1)+1}^{(2m+1)(N-2)}[2m+1]$ in the notation of \cite{Wang:2015mra,Xie:2016evu}.
We consider the principal nilpotent deformation of the flavor node on the right-hand side end. We find that there are indeed operators of fractional dimensions and rational central charges. 

This paper is organized as follows: In section \ref{sec:AD}, we review aspects of the generalized Argyres-Douglas theories we study in this paper. In section \ref{sec:SUquiver}, we consider $\CN=1$ deformations of certain $\CN=2$ $SU$ quiver gauge theories that flow to the generalized AD theories of type $A_{k-1}$. In section \ref{sec:SPSOquiver}, we consider the deformation of $Sp-SO$ gauge theories that, we conjecture, will flow to AD theories. We conclude in section \ref{sec:conclusion} with some remarks. \\

\emph{Note added:} While we were finishing this paper, we became aware that the $SU$-quiver gauge theories (\ref{eq:SUquiver}) that flow to the $(A_{k-1}, A_{Nk-1})$ and $(I_{m, mk}, S)$ AD theories were also independently discovered by \cite{Benvenuti:2017bpg} and reported earlier by one of the authors. We thank them for sharing this information. We coordinated the submission of this paper with them. 

\section{Review on generalized AD theories} \label{sec:AD}

A large class of 4d $\CN=2$ superconformal theories can be constructed by compactifying 6d $\CN=(2, 0)$ theory of type $\G \in ADE$ on a Riemann surface with a partial topological twist \cite{Gaiotto:2009hg,Gaiotto:2009we}. Such 4d theories are said to be in class $\CS$. One can construct generalized Argyres-Douglas theories by choosing the Riemann surface as a sphere with one irregular puncture \cite{Xie:2012hs,Wang:2015mra,Xie:2017vaf}. On top of this, it is possible to add one regular puncture. A regular puncture is labeled by an $SU(2)$ embedding into $\G$. When $\G=SU(N)$, the $SU(2)$ embeddings are in one-to-one correspondence with the partitions of $N$. The type of singularity determines the Seiberg-Witten curve. For example, if we choose $\G=A_{k-1}$ with the Riemann surface being a sphere with one irregular puncture of type $I_{k, N}$ (following the notation of \cite{Xie:2012hs}), we obtain the SW curve at the conformal phase as
\begin{align}
 x^k = z^N \ , 
\end{align}
with Seiberg-Witten 1-form given as 
\begin{align} \label{eq:SWformAkAn}
 \lambda_{SW} = x dz \ .
\end{align}
The SW differential should have scaling dimension 1, since the mass of the BPS particle is given as  $M = |Z| = |\int \lambda_{SW}|$. 
The AD theory has chiral operators parametrizing the Coulomb branch. They appear as the deformations of the singular curve. In this case, it can be deformed to give
\begin{align} \label{eq:curveAkAn}
 x^k = z^N + \sum_{\ell=2}^{k} \sum_{i=n+2 +\lfloor j(\ell-1)/k \rfloor }^{\ell n } u_{\ell, i} x^{k-\ell} z^{k n - i} \ , 
\end{align}
where $N=k n - j$ with $k, n, j \in \IZ_{\ge 0}$. 
The parameters $u_{\ell, i}$ appear in pairs $(u, v)$ so that $[u]+[v]=2$. The parameter that has scaling dimension greater than 1 is identified as the chiral operator, and the other is identified as the coupling constant appearing in the Lagrangian \cite{Argyres:1995xn}. 

Another way of constructing a class of AD theories is to start with type IIB superstring theory and compactify it on a Calabi-Yau 3-fold singularity of the form
\begin{align}
 W(x, y, z, w) = W_G (x, y) + W_{G'} (z, w) = 0 \ , 
\end{align}
with $(x, y, z, w) \in \IC^4$. Here $W_G$ are the equations governing the form of the singularity of ADE type and are given as follows:
\begin{align} 
\begin{split}
 W_{A_{n}} (x, y) &= x^{n+1} + y^2 \\
 W_{D_n} (x, y) &= x^{n-1} + x y^2 \\
 W_{E_6}(x, y) &= x^3 + y^4 \\
 W_{E_7} (x, y) & = x^3 + x y^3 \\
 W_{E_8} (x, y) &= x^3 + y^5  
\end{split}
\end{align}
The 4d $\CN=2$ superconformal theory obtained in this way is called the $(G, G')$ theory \cite{Cecotti:2010fi}. The Seiberg-Witten 1-form is replaced by a holomorphic 3-form given as
\begin{align}
 \Omega = \frac{dx\wedge dy \wedge dz \wedge dw}{dW} \ , 
\end{align}
and the mass of a BPS particle is given by \cite{Shapere:1999xr}
\begin{align}
 M_C = \int_C \Omega \ , 
\end{align}
where $C$ is a supersymmetric 3-cycle in the Calabi-Yau 3-fold. 

The chiral operator content of the theory is determined by the deformations of the singularity \cite{Shapere:1999xr}. Consider the ring of holomorphic functions of four variables $\IC[x, y, z, w]$, and take a quotient by the ideal generated by $dW$: 
\begin{align}
 \CR = \IC[x, y, z, w]/dW \,.
\end{align}
We can write an element in $\CR$ as $x^\a = x^i y^j z^k w^l$ modulo $dW = 0$. Then consider the deformation
\begin{align}
 W (x, y, z, w) \to W(x, y, z, w) + \sum_{x^\a \in \CR} u_{\a} x^\a \ . 
\end{align}
Combining this with the fact that $[\Omega]=1$, one can compute the scaling dimensions of the deformation parameters $u_\a$. Among them, the ones that have dimension greater than 1 are identified as the Coulomb branch operators.\footnote{There can be also mass parameters that have dimensions greater than or equal to 1. One can distinguish them by noticing that they do not pair up with other parameters so that the dimensions add up to 2.} 

One noticeable feature of the generalized AD theories is that quite often they admit exactly marginal deformations. This can be seen if some of the Coulomb branch operators have scaling dimension $2$, which includes exactly marginal operator in the same multiplet. It is widely believed that any exactly marginal deformation of an $\CN=2$ SCFT arises through a gauge interaction. Therefore it should be possible to take the extremely weak limit of the gauge couplings and decompose the theory into smaller AD theories with global symmetry. When there is a gauge coupling, there may be a dual description. The dual descriptions for the AD theories have been studied in \cite{Buican:2014hfa,Xie:2016uqq,Xie:2017vaf,Buican:2017fiq}.

\paragraph{$(A_{k-1}, A_{N-1})$ theory}
This theory can also be obtained in class $\CS$ by choosing $\G=A_{k-1}$ and the Riemann surface to be a sphere with one irregular puncture of type $I_{k, N}$. The Seiberg-Witten curve and the 1-form for this theory is given as in \eqref{eq:curveAkAn} and \eqref{eq:SWformAkAn}. 
From this curve and 1-form, one can deduce the scaling dimensions of the Coulomb branch operators of this theory: in fact, by noticing that
\begin{align}
 [x] = \frac{N}{N+k} \ , \quad [z]=\frac{k}{N+k}  \ , 
\end{align}
we obtain
\begin{align}
 [u_{\ell, i}] = \frac{ik - \ell j}{N+k} = \frac{ik - \ell j}{k+kn-j}\ . 
\end{align}
Among the $u_{i, j}$'s, the ones that have scaling dimension greater than $1$ are identified as the Coulomb branch operators. 

From the Coulomb branch operator spectrum and the curve, it is possible to compute the central charges of the theory \cite{Shapere:2008zf}. First, there is a relation\footnote{This relation is modified when an $\CN=2$ SCFT is obtained via gauging a discrete subgroup of the $U(1)_r$ symmetry \cite{Aharony:2016kai, Argyres:2016yzz}. }
\begin{align} \label{eq:STformula}
 2a - c = \frac{1}{4} \sum_i \left(2[u_i] - 1 \right) \ , 
\end{align}
where the sum is over the Coulomb branch operators. Another relation is
\begin{align}
 a = \frac{1}{4}R(A) + \frac{1}{6} R(B) +\frac{5r}{24} \ , \quad c = \frac{1}{3}R(B) + \frac{r}{6} \ , 
\end{align} 
where $r$ is the dimension of the Coulomb branch (sometimes called the rank of an $\CN=2$ SCFT) and 
\begin{align}
 R(A) = \sum_i [u_i] - r \ , 
\end{align}
while $R(B)$ is a quantity that can be computed from the SW curve. 

Now, coming back to the case of $(A_{k-1}, A_{N-1})$ theory, one can compute the central charges from the above relations. The function $R(B)$ is given as \cite{Xie:2013jc}
\begin{align}
 R(B) = \frac{(k-1)Nk(N-1)}{4(N+k)} \ . 
\end{align}
From this formula, we can compute the central charges for arbitrary $k$ and $N$. The formula simplifies when $N=nk$, with $n$ being an integer:
\begin{align} \label{eq:acAkAn}
 a= \frac{(k-1)(2k^2 n^2 + 2kn^2 - 5n-5)}{24(n+1)} \ , \quad c = \frac{(k-1)(k^2 n^2 + kn^2 - 2n-2)}{12(n+1)} \,.
\end{align}
The Coulomb branch operators are given as
\begin{align}
  [u_{\ell, i}] = \frac{i}{n+1}\ ,
\end{align}
with $\ell = 2, 3, \ldots, k$ and $i=n+2, n+3, \ldots, \ell n$. 

\paragraph{$(I_{k, N}, Y)$ theory}
One can engineer this type of AD theory by compactifying 6d $\CN=(2, 0)$ theory of type $\G = A_{k-1}$ on a sphere with an irregular puncture of type $I_{k, n}$ and a regular puncture of type $Y$. This theory has a global symmetry $SU(k)$ if $(k, N)=1$ and $SU(k)\times U(1)^{k-1}$ if $N$ is divisible by $k$. Here $Y$ labels the partitions of $k$. When $Y=[1, \ldots, 1]$ is the full puncture, the SW curve will be of the form
\begin{align}
 x^k + z^N + \frac{m_k}{z^k} + \ldots =  0  \ , 
\end{align}
where $m_k$ is one of the mass parameters of $SU(k)$. 
Here we omitted the deformations. 

When $N=kn$, the central charges are given as
\begin{align}
 a =  \frac{1}{48}(k-1)(4k^2 n +4k^2 + 4kn - k - 10)\ , \quad c = \frac{1}{12}(k-1)(k^2 n+k^2+kn -2) \ ,
\end{align}
and the dimensions of the Coulomb branch operators are
\begin{align}
 [u_{\ell, i}] = \frac{2i}{n+1} \qquad (\ell=2,3, \ldots, k,~~ i=n+2, \ldots, \ell (n+1)-1) \ . 
\end{align}
When $Y=S \equiv [k-1, 1]$ is the simple puncture and $N=kn$, the central charges are
\begin{align} \label{eq:acIY}
  a = \frac{k(2k^2 n^2 + 6kn - 2n^2 - 5n+1)}{24(n+1)} \ , \quad
  c=\frac{k(k^2 n^2 + 3kn - n^2 - 2n+1)}{12(n+1)} \ , 
\end{align}
and the dimensions of the Coulomb branch operators are
\begin{align}
 [u_{\ell, i}] = \frac{2i}{n+1}  \, ,
\end{align}
with $\ell=2,3, \ldots, k$ and $i=n+2, n+3, \ldots, \ell n+1$.

\paragraph{$(A_k, D_n)$ theory}
Some of the AD theories do not have a class $\CS$ realization. One example is the $(A_k, D_n)$ theory with $k>1$. This can be easily realized from type IIB string theory compactified on the Calabi-Yau 3-fold singularity 
\begin{align}
x^{k+1} + z^{n-1} + z y^2 + w^2 = 0 \ . 
\end{align}
Since the holomorphic 3-form has scaling dimension 1, we can deduce the scaling dimensions of the coordinates to be
\begin{align}
[x]=\frac{2 (n-1)}{k+2 n-1}, ~~ [y]=\frac{(k+1) (n-2)}{k+2 n-1}, 
~~ [z]=\frac{2 (k+1)}{k+2 n-1}, ~~ [w]\ = \frac{(k+1) (n-1)}{k+2 n-1} . 
\end{align}
This allows us to compute the scaling dimension of the Coulomb branch operators that can be obtained as a deformation of the singularity. 

For example, the $(A_3, D_5)$ theory has Coulomb branch operators of dimension 
\begin{align}
\label{eq:A3D5ColBr}
 \Delta(u) \in \left\{ \frac{4}{3}, \frac{4}{3}, \frac{4}{3}, \frac{5}{3}, 2, 2, \frac{8}{3} \right\} \ . 
\end{align}

We can also compute the central charges of this theory. For the $(G, G')$ theory, the BPS quiver, that encodes the massive BPS spectrum in the Coulomb branch, has the form of a product of the Dynkin diagrams for $G$ and $G'$. From this information, one can form quantum monodromy operator $\CM$ \cite{Cecotti:2010fi}. The trace of monodromy $\tr \CM^N$ (or its power) can be associated to a 2d chiral algebra where the 2d central charge $c_{2d}$ is given in terms of 4d central charge $c_{4d}$ ad $c_{2d} = 12N c_{4d}$. At the same time, one can compute the (effective) central charge by studying the scaling behavior of $\tr \CM$ \cite{Cecotti:2013lda, Cecotti:2015lab}. To this end, we get
\begin{align}
 c = \frac{1}{12} \left( \frac{r_G r_{G'} h_G h_{G'}}{h_G + h_{G'}} + 2 r \right) \ , 
\end{align}
where $r_G$, $h_G$ are the rank and the dual Coxeter number of $G$ respectively, and $r$ is the dimension of the Coulomb branch (rank) of the theory. Once we know the $c$ and the Coulomb branch spectrum, we can deduce the other central charge $a$ from the relation \eqref{eq:STformula}.

\section{AD theories from quivers with $SU(n)$ gauge groups}
\label{sec:SUquiver}

\subsection{Lagrangian for $(A_{m-1},A_{N m-1})$ theory}
Given the success of $\CN=1$ nilpotent deformations in producing effective Lagrangians that flow to Argyres-Douglas theories of type $(A_1, G)$ with $G=A_n, D_n$, it is natural to wonder if similar deformations of quiver theories can lead to more general AD theories  such as $(A_k, G)$. Motivated by this, we consider the following $4d$ $\CN=2$ quiver gauge theory. Start with a theory having an $SU(N)$ gauge symmetry and $2N$ hypermultiplets in the fundamental representation. As is well known, this theory is superconformal and has an $SU(2N)$ flavor symmetry which we wish to gauge. One way to maintain superconformality will then be to add to this theory, $3N$ hypermultiplets transforming in the fundamental representation of the newly minted $SU(2N)$ gauge group. The resulting theory will have $SU(3N)$ flavor symmetry which we can again gauge if we wish and add $4N$ hypermultiplets transforming in the fundamental representation of the $SU(3N)$ gauge group. As is obvious, this process can be continued indefinitely to produce a quiver of any desired length. A generic quiver of this kind will therefore consists of an $SU(N) \times SU(2N) \times \cdots \times SU(mN-N)$ gauge group and an $SU(m N)$ flavor symmetry group, and is given as in the following quiver diagram:
\begin{align}
 (N)-(2N)-(3N)-\ldots-(mN-N)-\boxed{mN}
\end{align}

Let $\phi_\ell$ $(\ell= 1,\hdots,m-1)$ denote the adjoint chiral multiplet associated to the $\CN=2$ vector multiplet for the $SU(lN)$ gauge node in the above quiver. Also, let $(Q_\ell, \widetilde{Q}_\ell)$ be the chiral multiplets forming the hypermultiplet that transforms in the bifundamental representation of $SU(\ell N) \times SU(\ell N+N)$. We will use $\mu_\ell$ to denote the $SU(\ell N)$ moment map operator formed from $(Q_\ell, \widetilde{Q}_\ell)$ while $\widetilde{\mu}_\ell$ will denote the $SU(\ell N+N)$ moment map operator formed from $(Q_\ell, \widetilde{Q}_\ell)$. The superpotential of the above quiver is then given by
\be
\label{eq:SupPotSUquiver}
W_{\CN=2} = {\rm Tr}\phi_1 \mu_1 + \sum_{\ell=2}^{m-1} {\rm Tr} \phi_\ell (\mu_\ell -\widetilde{\mu}_{\ell-1}) \ .
\ee 
We now introduce a chiral multiplet $M$ transforming in the adjoint representation of the $SU(m N)$ flavor symmetry, switch on a superpotential term given by
\be
\label{eq:N1defSU}
\delta W = {\rm Tr}\widetilde{\mu}_{m-1} M \ ,
\ee 
and turn on a nilpotent vev $\rho: SU(2) \hookrightarrow SU(m N)$ for $M$. This will break the $SU(m N)$ flavor symmetry down to the commutant of the $SU(2)$ embedding specified by $\rho$. The supermultiplets containing the Goldstone modes corresponding to the broken flavor symmetry generators will decouple in the IR. These can be easily identified using the arguments delineated in \cite{Gadde:2013fma}. The chiral multiplets that become massive as a result of the above vev can also be easily integrated out, resulting in a ``Fan'' \cite{Agarwal:2014rua}. As was explained in \cite{Agarwal:2014rua}, this theory has two candidate $R$-symmetries which we call $J_+$ and $J_-$. We assume that the $\CN=1$ theory so obtained flows to a fixed point in the IR. The $R$-symmetry of the superconformal fixed point is then given by a linear combination of $J_+$ and $J_-$. This correct linear combination can be obtained using the technique of $a$-maximization \cite{Intriligator:2003jj}.  

The various possibilities for the vev $\rho$ are classified by partitions of $N m$. Upon scanning through the space of all such possibilities, we reached the conclusion that the most interesting of these is when $\rho$ corresponds to the principal embedding which sends the fundamental representation of $SU(m N)$ into the $m N$ dimensional irrep of $SU(2)$. For other choices of $\rho$, we generically find irrational central charges, implying that they do not exhibit supersymmetry enhancement in the IR. In the rest of this section, we will therefore mostly focus on the principal embedding case. 

For $\vev{M} = \rho_{principal}$, the $SU(m N)$ flavor symmetry is completely broken. Recall that before giving a vev, there were $N m$ quarks transforming in the fundamental representation of the $SU(mN- N)$ gauge symmetry. Due to the vev $\vev{M}$, $N m-1$ of these will become massive and get integrated out. The fields forming the rest of the quiver will not be affected.\footnote{However, as we discuss later, some of the fields and operators will hit the unitarity bound and decouple along the RG flow.}   The matter content of the theory obtained after integrating out the massive fields and removing the decoupled modes, is given in table \ref{tab:SUPrin}.
 \begin{table}[h]
	\centering
	\begin{tabular}{|c|c|c|c|c|c|c|c|}
		\hline
		fields & $SU(\ell N)$ &$SU(\ell N+N)$ & $SU(mN-N)$ &$U(1)_l$& $U(1)_{m-1}$& $J_+$ & $J_-$\\
		\hline \hline
		$Q_\ell$ & $\ytableausetup{smalltableaux}\ydiagram{1}$ & $\overline{\ytableausetup{smalltableaux}\ydiagram{1}}$ & $\mathbf{1}$ & 1& 0&1&0 \\ \hline
		$\widetilde{Q}_\ell$ & $\overline{\ytableausetup{smalltableaux}\ydiagram{1}}$ & $\ytableausetup{smalltableaux}\ydiagram{1}$ & $\mathbf{1}$ & -1& 0&1&0 \\ \hline
		$q$ & $\mathbf{1}$ & $\mathbf{1}$ &$\ytableausetup{smalltableaux}\ydiagram{1}$ &0&1& 1 & $1- N m$ \\ \hline
		$\widetilde{q}$ & $\mathbf{1}$ & $\mathbf{1}$ &$\overline{\ytableausetup{smalltableaux}\ydiagram{1}}$ & 0&-1&1 & $1- N m$ \\ \hline
		$\phi_\ell$ & \rm{adj} & $\mathbf{1}$ & $\mathbf{1}$ &0&0& 0 & 2 \\ \hline
		$\phi_{m-1}$ &$\mathbf{1}$ & $\mathbf{1}$ &\rm{adj} &0&0& 0 & 2 \\ \hline
		$M_j$ & $\mathbf{1}$ &  $\mathbf{1}$ &  $\mathbf{1}$ & 0&0&0 & $2j+2$ \\ \hline
	\end{tabular}
	\caption{Matter content for the ``Lagrangian description'' of the $(A_{m-1}, A_{N m-1})$ theory. Here $\ell$ runs from $1$ to $m-2$ and $j$ for the $M_j$ runs from $1, \ldots, N m-1$. $U(1)_\ell$ is a baryonic global symmetry acting on $(Q_\ell, \widetilde{Q}_\ell)$ . } 
	\label{tab:SUPrin}
\end{table}	 
The superpotential of this system can easily be written down using the results of \cite{Agarwal:2014rua}. The upshot is to write all possible combination of the fields so that each term in the superpotential has the charge $(J_+, J_-)=(2, 2)$. 

The IR $R$-charge is given in terms of $J_+$ and $J_-$ by 
\be
R_{\CN=1} = \frac{1+\epsilon}{2} J_+ + \frac{1-\epsilon}{2}J_- \ ,
\ee 
where $\epsilon$ is determined via $a$-maximization. For the theory being described here, the trial central charges $a(\epsilon)$ and $c(\epsilon)$ are given by
\be
\label{eq:SUafun}
\begin{split}
	a(\epsilon)&=\frac{3}{16}-\frac{3 m}{16}+\frac{3 m N}{64}-\frac{3 m N^2}{64}+\frac{9 m^3 N^2}{128}-\frac{9 m^3 N^3}{64}+\frac{9 m^3 N^4}{128}\\
	&~~+\left(-\frac{3 m}{32}-\frac{3 m N}{32}+\frac{3 m N^2}{16}+\frac{9
		m^3 N^2}{128}+\frac{9 m^3 N^3}{64}-\frac{27 m^3 N^4}{128}\right) \epsilon\\
	&~~+\left(-\frac{9 m N}{64}-\frac{9 m N^2}{64}-\frac{9 m^3 N^2}{128}+\frac{9 m^3 N^3}{64}+\frac{27 m^3 N^4}{128}\right)
	\epsilon ^2\\
	&~~+\left(\frac{9 m}{32}-\frac{9 m^3 N^2}{128}-\frac{9 m^3 N^3}{64}-\frac{9 m^3 N^4}{128}\right) \epsilon ^3 \ ,
\end{split}	
\ee 
and
\be
\label{eq:SUcfun}
\begin{split}
	c(\epsilon)&=\frac{1}{8}-\frac{m}{8}+\frac{5 m N}{64}-\frac{5 m N^2}{64}+\frac{9 m^3 N^2}{128}-\frac{9 m^3 N^3}{64}+\frac{9 m^3 N^4}{128}\\
	&~~+\left(-\frac{5 m}{32}-\frac{m N}{16}+\frac{7 m N^2}{32}+\frac{9 m^3
		N^2}{128}+\frac{9 m^3 N^3}{64}-\frac{27 m^3 N^4}{128}\right) \epsilon \\
	&~~+\left(-\frac{9 m N}{64}-\frac{9 m N^2}{64}-\frac{9 m^3 N^2}{128}+\frac{9 m^3 N^3}{64}+\frac{27 m^3 N^4}{128}\right)
	\epsilon ^2\\
	&~~+\left(\frac{9 m}{32}-\frac{9 m^3 N^2}{128}-\frac{9 m^3 N^3}{64}-\frac{9 m^3 N^4}{128}\right) \epsilon ^3 \ .
\end{split}
\ee
Upon maximizing $a(\epsilon)$, we find that the IR $R$-charges and hence the IR dimensions of the various operators so obtained are such that some of the gauge singlets and gauge invariant operators in the theory decouple since they violate the unitarity bound $R\ge \frac{2}{3}$. Hence, these have to be removed from the interacting theory, giving us a corrected $a$-function \cite{Kutasov:2003iy}.\footnote{Also see \cite{Morita:2011cs,Agarwal:2012wd,Safdi:2012re} for a similar discussion about subtleties in 3d theories arising due to decoupling of gauge invariant operators.} In practice, what happens is that we have to repeat the above cycle of $a$-maximizing and checking for possible decoupling of operators multiple times, until we reach a stage when no gauge invariant fields/operators decouple any more. 

For the quivers at hand, by explicitly computing for some low lying values of $m$ and $N$, we found that at the end of the RG flow, the following operators decouple: ${\rm Tr} \phi^k_\ell \ , \forall \, 1 \leq \ell \leq m-1 \ \text{and} \  2\leq k \leq \min(N+1,\ell N)$. Along with these the gauge singlet fields $M_j$, $\forall \ 1 \leq j \leq N$ also decouple. Removing these from the interacting theory implies that the corrected $a$-function that describes the IR fixed point is given by
\be
\label{eq:correctedASU}
\begin{split}
a_{corr}(\epsilon)&=\frac{3}{128} (m-1) \left(-8+4 N+3 m (1+m) N^2-6 \left(2+m+m^2\right) N^3+3 m (1+m) N^4\right)\\
&~~-\frac{3}{128}(m-1) (1+3 N) \left(4-8 N-3 \left(4+m+m^2\right) N^2+3 m (1+m) N^3\right) \epsilon\\
&~~+\frac{9}{128} (m-1) N (1+N) \left(-12-\left(12+m+m^2\right) N+3 m (1+m) N^2\right) \epsilon ^2\\
&~~-\frac{9}{128} (m-1) (1+N)^2 \left(-4-4 N+m (1+m) N^2\right) \epsilon ^3 \ ,
\end{split}
\ee 
while the corrected $c$-function is given by
\be
\label{eq:correctedCSU}
\begin{split}
	c_{corr}(\epsilon)&=\frac{1}{128} (m-1) \left(-16+20 N+9 m (1+m) N^2-18 \left(2+m+m^2\right) N^3+9 m (1+m) N^4\right)\\
	&~~-\frac{1}{128} (m-1) \left(20+20 N-9 \left(12+m+m^2\right) N^2-18 \left(6+m+m^2\right) N^3+27
	m (1+m) N^4\right) \epsilon\\
	&~~+\frac{9}{128} (m-1) N (1+N) \left(-12-\left(12+m+m^2\right) N+3 m (1+m) N^2\right) \epsilon ^2\\
	&~~-\frac{9}{128} (m-1) (1+N)^2 \left(-4-4 N+m (1+m) N^2\right)
	\epsilon ^3 \ .
\end{split}
\ee 
The $a$-function given in \eqref{eq:correctedASU} maximizes at $\epsilon = \frac{1+3 N}{3 (1+N)}$. Substituting this back we find that the IR central charges of the interacting theory are given by
\be
\label{eq:ASUint}
\begin{split}
	a_{IR} &= \frac{(m-1) \left(-5-5 N+2 m (1+m) N^2\right)}{24 (1+N)} \ ,\\
	c_{IR} &= \frac{(m-1) \left(-2-2 N+m (1+m) N^2\right)}{12 (1+N)} \ .
\end{split}
\ee
These values match exactly with the central charges of the $(A_{m-1}, A_{N m-1})$ type Argyres-Douglas theory given in \eqref{eq:acAkAn}. We therefore are tempted to conjecture that the above set of theories experience SUSY enhancement and flow to the fixed points described by the $(A_{m-1}, A_{N m-1})$ AD theories. 

A crucial piece of evidence that supports our conjecture comes from matching the spectrum of Coulomb branch operators of the afore-mentioned AD theories to the operator spectrum of the theories obtained through the nilpotent deformation being described here.\footnote{In fact there exist theories with same central charges $(a,c)$ but different Coulomb branch operator spectrum: for example both $(A_2,A_8)$ and $(A_1, D_{18})$ have $(a,c) = (\frac{49}{12},\frac{25}{6})$, but the dimensions of Coulomb branch operators are $(\frac{5}{4},\frac{5}{4},\frac{3}{2},\frac{3}{2},\frac{7}{4},2,\frac{9}{4})$ and $(\frac{10}{9},\frac{11}{9},\frac{4}{3},\frac{13}{9},\frac{14}{9},\frac{5}{3},\frac{16}{9},\frac{17}{9})$ respectively.}
 To see this notice that the Coulomb branch operators of the $(A_{m-1}, A_{N m-1})$ AD theory are given by 
the set $\{ u_{s,i}\ ; \ 2 \leq s \leq m \ , \ N+2\leq i \leq s N  \} $ with the dimension of $u_{s,i}$ being
\be
[u_{s,i}]=\frac{i}{N+1} \ .
\ee 
On the other hand in the Lagrangians being described here, the partial list of gauge invariant chiral operators is given by $\{{\rm Tr}\phi^k_\ell \ ; \ 2 \leq \ell \leq m-1 \ , \ N+2 \leq k \leq \ell N \} \cup \{M_j \ ; \ N+1 \leq j \leq N m-1 \}$. The dimension of these operators at the IR fixed point can be easily computed by using the relation $\Delta = \frac{3}{2} R_{IR}$, which gives
\be
[{\rm Tr}\phi^k_\ell] = \frac{k}{N+1} \ , \ [M_j] = \frac{j+1}{N+1} \,.
\ee 
We therefore see that $[{\rm Tr}\phi^k_\ell]$ and $u_{\ell,k}$ , $\forall \ 2 \leq \ell \leq m-2 \ , \ N+2 \leq k \leq \ell N$ are in one-to-one correspondence, while $M_j$ corresponds to $u_{m,j}$.

\subsection{Lagrangian for $(I_{m,mk},S)$ theory}

Another class of $4d$ $\CN=2$ superconformal quivers that proved to show interesting behaviour upon nilpotent deformation is given as follows. The quiver consists of $m-1$ gauge nodes, with the $\ell$-th node carrying an $SU(k \ell +1)$ gauge symmetry. There are bifundamental hypermultiplets connecting the $\ell$-th node to the $(\ell+1)$-th node. This causes the $\beta$-function for the gauge coupling at each node, except the first and the last one, to vanish. In order to make the $\beta$-function at the first and the last node to vanish, we then need to couple one more fundamental hyper at the first node and $mk+1$ fundamental hypermultiplets at the last node. We can represent this theory using the quiver diagram as follows:
\begin{align}
 \boxed{1}-(k+1)-(2k+1)-\ldots-(mk-k+1)-\boxed{mk+1}
\end{align}
This quiver gauge theory has an $SU(mk+1) \times U(1)^{m} $ flavor symmetry. 

We now deform this quiver by coupling a chiral multiplet $M$ transforming in the adjoint representation of $SU(m k+1)$ via the superpotential term
\be
\delta W = {\rm Tr} \widetilde{\mu} M \ ,
\ee    
where $\widetilde{\mu}$ is the moment map operator for the $SU(mk+1)$ flavor symmetry in the undeformed quiver. Upon studying the outcome of giving a nilpotent vev $\rho:SU(2)\hookrightarrow SU(mk+1)$ to $M$, we arrive at the conclusion that in this case also, the vev corresponding to the principal embedding is the most interesting. The other choices generically give irrational central charges at the end of the RG flow, therefore no SUSY enhancement occurs. We thus focus on the principal embedding in the rest of this discussion. 

The matter content of the theory obtained  via this deformation (upon integrating out the massive quarks along with removing the decoupled fields) is given in table \ref{tab:SUPrinQ2} . 
 \begin{table}[t]
	\centering
	\begin{tabular}{|c|c|c|c|c|c|c|c|}
		\hline
		fields &$SU(k+1)$& $SU(k\ell+1)$ &$SU( (\ell+1)k+1)$ & $SU((m-1)k+1)$ & $J_+$ & $J_-$\\
		\hline \hline
		$q_1$ & $\ytableausetup{smalltableaux}\ydiagram{1}$ & $\mathbf{1}$ & $\mathbf{1}$&$\mathbf{1}$ & 1&0 \\ \hline
		$\widetilde{q}_1$ & $\overline{\ytableausetup{smalltableaux}\ydiagram{1}}$ & $\mathbf{1}$ & $\mathbf{1}$&$\mathbf{1}$ & 1&0 \\ \hline
		$Q_\ell$ &$\mathbf{1}$ &$\ytableausetup{smalltableaux}\ydiagram{1}$ & $\overline{\ytableausetup{smalltableaux}\ydiagram{1}}$ & $\mathbf{1}$ & 1&0 \\ \hline
		$\widetilde{Q}_\ell$ &$\mathbf{1}$ &$\overline{\ytableausetup{smalltableaux}\ydiagram{1}}$ & $\ytableausetup{smalltableaux}\ydiagram{1}$ & $\mathbf{1}$ &1&0 \\ \hline
		$q_m$ & $\mathbf{1}$ &$\mathbf{1}$ & $\mathbf{1}$ &$\ytableausetup{smalltableaux}\ydiagram{1}$ & 1 & $-mk$ \\ \hline
		$\widetilde{q}_m$ &$\mathbf{1}$& $\mathbf{1}$ & $\mathbf{1}$ &$\overline{\ytableausetup{smalltableaux}\ydiagram{1}}$ & 1 & $-mk$ \\ \hline
		$\phi_\ell$ &$\mathbf{1}$ &\rm{adj} & $\mathbf{1}$ & $\mathbf{1}$ & 0 & 2 \\ \hline
		$M_j$ & $\mathbf{1}$ &  $\mathbf{1}$ &  $\mathbf{1}$ & $\mathbf{1}$&0 & $2j+2$ \\ \hline
	\end{tabular}
	\caption{Matter content for the ``Lagrangian description'' of the $(I_{m,m k},S)$ theory. Here $\ell$ runs from $1$ to $m-2$, while the index $j$ in $M_j$ runs from 1 to $m k$. For the sake of brevity, we have not included the flavor symmetries acting on the various bifundamental hypermultiplets. } 
	\label{tab:SUPrinQ2}
\end{table}	
The broad picture of what follows is same as the discussion in the previous subsection. The details are of course different. The trial $a$ and the $c$-functions of this Lagrangian gauge theory are given by
\begin{align}
\label{eq:SUafun2}
\begin{split}
	a(\epsilon)&=\frac{3}{128} m \left(1+3 k^3 (3-2 m) m+3 k^4 m^2+k (-16+9 m)+k^2 \left(7-18 m+3 m^2\right)\right)\\
	&~~+\frac{3}{128} (1-k) m \left(5+9 k^3 m^2+3 k^2 m (9+m)+k (19+9 m)\right) \epsilon\\
	&~~+\frac{9}{128}
	(1+k) m \left(-3+k (7-3 m)-k^2 (-9+m) m+3 k^3 m^2\right) \epsilon ^2\\
	&~~-\frac{9}{128} m \left(-1+k^4 m^2+3 k (2+m)+k^3 m (3+2 m)+k^2 \left(3+6 m+m^2\right)\right) \epsilon ^3 \ ,
\end{split}
\end{align}
and
\begin{align}
\label{eq:SUcfun2}
\begin{split}
	c(\epsilon)&=\frac{1}{128} m \left(11+9 k^4 m^2-9 k^3 m (2 m-3)+k (27 m-44)+k^2 \left(17-54 m+9 m^2\right)\right)\\
	&~~+\frac{1}{128} (1-k) m \left(7+27 k^3 m^2+9 k^2 m (9+m)+k (53+27 m)\right) \epsilon
	\\
	&~~+\frac{9}{128} (1+k) m \left(-3+k (7-3 m)-k^2 (-9+m) m+3 k^3 m^2\right) \epsilon ^2\\
	&~~-\frac{9}{128} m \left(-1+k^4 m^2+3 k (2+m)+k^3 m (3+2 m)+k^2 \left(3+6 m+m^2\right)\right) \epsilon ^3 \ .
\end{split}
\end{align}
Repeating the cycle of $a$-maximizing and removing gauge invariant operators and fields that hit the unitarity bound, we find that all of ${\rm Tr}\phi^i_\ell \ , \ \forall \ 1 \leq \ell \leq m-1 \ , 2 \leq i \leq k+1 $, hit the unitarity bound and decouple as free fields. At the same time, the gauge singlet fields $M_j \ , \ 1 \leq j \leq k$, also decouple. The $a$ and the $c$-function describing the flow of the interacting theory (after removing the free fields that decouple) are then given by
\be
\label{eq:SUAcor2}
\begin{split}
	a_{corr}(\epsilon)&=\frac{3}{128} m \left(1+k (9 m-14)+k^3 \left(9 m-6 m^2 -6 \right)+3 k^4 \left(m^2 -1 \right)+3 k^2 \left(2-6 m+m^2\right)\right)\\
	&~~- \frac{3}{128} (1+3 k) m \left(-5+k-9 k m+3 k^3
	\left(m^2 -1\right)-3 k^2 \left(3-3 m+m^2\right)\right) \epsilon\\
	&~~+\frac{9}{128} (1+k) m \left(-3-3 k (1+m)+3 k^3 \left(m^2 -1 \right)-k^2 \left(11-9 m+m^2\right)\right) \epsilon
	^2\\
	&~~-\frac{9}{128} (1+k)^2 m \left(-1+k (-4+3 m)+k^2 \left(m^2 -1\right)\right) \epsilon ^3\ ,
\end{split}
\ee  
and
\be
\label{eq:SUCcor2}
\begin{split}
	c_{corr}(\epsilon)&=\frac{1}{128} m \left(11+k (27 m-34)+9 k^4 \left(m^2 -1\right)+9 k^2 \left(2-6 m+m^2\right)-9 k^3 \left(2-3 m+2 m^2\right)\right)\\
	&~~+\frac{1}{128} m \left(7+k (34+27 m)-27 k^4 \left(m^2 -1\right)+9
	k^2 \left(2+6 m+m^2\right)+9 k^3 \left(10-9 m+2 m^2\right)\right) \epsilon\\
	&~~+\frac{9}{128} (1+k) m \left(-3-3 k (1+m)+3 k^3 \left(m^2 -1\right)-k^2 \left(11-9 m+m^2\right)\right) \epsilon
	^2\\
	&~~-\frac{9}{128} (1+k)^2 m \left(-1+k (-4+3 m)+k^2 \left(m^2 -1\right)\right) \epsilon ^3 \ .
\end{split}
\ee 
The $a$-function given in \eqref{eq:SUAcor2} maximizes at $\epsilon = \frac{3 k+1}{3k+3}$. We thus find that the IR central charges of the interacting theory are given by
\be
\begin{split}
	a_{IR} &= \frac{m \left(1+k (6 m-5)+2 k^2 \left(m^2-1\right)\right)}{24 (1+k)}\ , \\
	c_{IR} &= \frac{m \left(1+k (3 m-2)+k^2 \left(m^2-1\right)\right)}{12 (1+k)} \ .
\end{split}
\ee 
These match perfectly with the central charges of the $(I_{k,m k},S)$ type AD theory given in \eqref{eq:acIY}. 

Let us now compare the operator spectrum of the two theories. The Coulomb branch operators of the $(I_{k,m k},S)$ theory are given by $\{ u_{\ell,i} \ ; \ 2 \leq \ell \leq m \ , \ k+2 \leq i \leq \ell k+1 \}$. Their respective dimensions are
\be
[u_{\ell,i}] = \frac{i}{k+1} \ .
\ee 
Meanwhile, the IR spectrum of chiral operators of the Lagrangian described here contains the following operators: $\{ {\rm Tr}\phi^{i}_\ell \ , \ 2 \leq \ell \leq m-1 \ , \ k+2 \leq i \leq \ell k +1  \} \cup \{M_j \ ,\ k+1 \leq j \leq m k \}$. Their respective IR dimensions are 
\be
[ {\rm Tr}\phi^{k}_\ell]=\frac{k}{k+1} \ , \ [M_j] = \frac{j+1}{k+1} \ .
\ee
It is straightforward to see that ${\rm Tr}\phi^{i}_\ell$ corresponds to $u_{\ell,i}$ while $M_j$ corresponds to $u_{m,j+1}$. This uncanny match of the spectra in the two theories provides further credence to our conjecture.

\section{AD theories from quivers with $SO/Sp$ gauge groups}
\label{sec:SPSOquiver}
In addition to quivers with unitary gauge groups, we can also consider 4d $\CN=2$ quivers with symplectic and orthogonal gauge groups. These too show some very interesting behaviour under $\CN=1$ nilpotent deformations. In the following, we  describe in detail our analysis of these quivers: they consist of alternating symplectic and orthogonal gauge groups with half-hypermultiplets transforming in the bi-fundamental representation of the gauge symmetry associated to any two consecutive pair of nodes.  



The first class of quivers we are interested in starts with a node with $Sp(N)$ gauge symmetry. The next node then carries $SO(4N+4)$ gauge symmetry. We can now continue this chain by making sure that we add enough fundamental hypermultiplets at each gauge node to make its $\beta$-function vanish. Depending on the type of the last gauge node, the flavor symmetry carried by the quiver will either be described by an orthogonal or a symplectic group.

Alternatively, we can choose the first node of the quiver to be $SO(N)$. The next node is then given by $Sp(N-2)$ and so on. Once again, depending upon the choice of last gauge node, the flavor symmetry of the quiver will be described by an orthogonal or a symplectic group.

We find that when the quivers carry a symplectic flavor group, the $\CN=1$ nilpotent deformations do not seem to show any interesting feature: the central charges are mostly irrational, which leads us to believe that these cannot have SUSY enhancement. Once in a while it happens that the central charges do become rational, however these cases do not seem to follow any fixed pattern, and neither were we able to match their central charges with those of any known $\CN=2$ theories: we conjecture that there is no SUSY enhancement in these cases. This is consistent with the conjectures of \cite{Agarwal:2016pjo}.   

However, it turns out that when the flavor group of the quiver is of $D_N = SO(2N)$ type, then the nilpotent deformation corresponding to the principle nilpotent orbit always gives an interacting IR theory with rational central charges.


\subsection{Lagrangian for $(A_{2m-1}, D_{2Nm+1})$ theory} 

Let us consider a quiver with $m$ gauge nodes carrying a symplectic gauge group and $m-1$ nodes carrying an orthogonal gauge group. The beginning gauge node of the quiver is chosen to be $Sp(N)$ while the $k$-th symplectic gauge node is given by $Sp(2k N-N)$. Meanwhile the $k$-th orthogonal gauge node carries $SO(4N k +2)$ gauge symmetry. The $\beta$-function at each gauge node vanishes except for the gauge node at the beginning of the quiver. To make the gauge coupling at this node marginal, we add 2 half-hypermultiplets transforming in the fundamental representation of the $Sp(N)$ gauge symmetry. Thus the total flavor symmetry of the quiver is given by $SO(2)\times SO(4mN+2)$. This quiver can be drawn as follows: 
\begin{align}
\label{eq:AnDkquiver}
\begin{split}
\boxed{SO(2)}-Sp(N)-SO(4N+2)-Sp(3N)-\ldots-Sp(2mN-N)-\boxed{SO(4mN+2)} 
\end{split}
\end{align}

As has been the theme of this paper, we now consider $\CN=1$ nilpotent deformations of this quiver by coupling a gauge singlet chiral superfield $M$. Here $M$ is chosen to transform in the adjoint representation of the $SO(4mN+2)$ flavor symmetry of the quiver. We now turn on vevs $\vev{M}=\rho$ given by nilpotent orbits of $SO(4mN+2)$, that is $\rho: SU(2)\hookrightarrow SO(4mN+2)$, and analyse the data at the $IR$ fixed point of this deformed quiver. Once again, the most interesting case turns out to be when $\rho$ corresponds to the principal nilpotent orbit of $SO(4mN+2)$. The matter content of the the Lagrangian obtained via this deformation, is summarized in table 
\ref{tab:AkDnLag}.
 \begin{table}[t]
	\centering
	\begin{tabular}{|c|c|c|c|c|c|c|c|}
		\hline
		fields& $Sp(2kN-N)$ &$SO(4Nk+2)$&$Sp(2kN+N)$ &$Sp(2mN-N)$&$J_+$ & $J_-$\\
		\hline \hline
		$Q_k$  &$\ytableausetup{smalltableaux}\ydiagram{1}$ & $\ytableausetup{smalltableaux}\ydiagram{1}$&$\mathbf{1}$ & $\mathbf{1}$ &1&0 \\ \hline
		$Q'_k$ &$\mathbf{1}$ &$\ytableausetup{smalltableaux}\ydiagram{1}$ & $\ytableausetup{smalltableaux}\ydiagram{1}$&$\mathbf{1}$  &1&0 \\ \hline
		$q'_m$ & $\mathbf{1}$ &$\mathbf{1}$ & $\mathbf{1}$ &$\ytableausetup{smalltableaux}\ydiagram{1}$ & 1 & $0$ \\ \hline
		$q_m$ &$\mathbf{1}$& $\mathbf{1}$ & $\mathbf{1}$ &$\ytableausetup{smalltableaux}\ydiagram{1}$ & 1 & $-4mN$ \\ \hline
		$\phi_{Sp,k}$  & $\mathbf{1}$ &\rm{adj} &$\mathbf{1}$ & $\mathbf{1}$ & 0 & 2 \\ \hline
		$\phi_{SO,k}$  &$\mathbf{1}$ & $\mathbf{1}$&\rm{adj}&$\mathbf{1}$ & 0 & 2 \\ \hline
		$M_{j=2\ell-1}$ & $\mathbf{1}$ &  $\mathbf{1}$ &  $\mathbf{1}$ & $\mathbf{1}$&0 & $2j+2$ \\ \hline
		$M'_{j=2mN}$ & $\mathbf{1}$ &  $\mathbf{1}$ &  $\mathbf{1}$ & $\mathbf{1}$&0 & $4mN+2$ \\ \hline
	\end{tabular}
	\caption{Matter content for the ``Lagrangian description'' of the $(A_{2m-1}, D_{2mN+1})$ theory. For the sake of brevity, we have not shown the $SO(2)$ flavor symmetry under which $q$ transforms as a doublet. Here $k$ runs from $0$ to $m-1$. At $k=0$, $SO(4Nk+2) \rightarrow SO(2)$, which is the flavor symmetry of the theory. The index $\ell$ in $M_{j=2\ell-1}$ runs from 1 to $2m N$.
	} 
	\label{tab:AkDnLag}
\end{table}
For this deformation, the central charges of the interacting theory (obtained after decoupling all operators and fields that hit the unitarity bound) at the fixed point are always rational and are given by the following functions of $m$ and $N$:
\be
\label{eq:centAkDn}
\begin{split}
	a_{IR} &= \frac{24 m^2 N+32 m^3 N^2-5 (1+2 N)+m \left(5-5 N-8 N^2\right)}{24+48 N} \ , \\ 
	c_{IR} &= \frac{-1-2 N+6 m^2 N+8 m^3 N^2-m \left(-1+N+2 N^2\right)}{6+12 N} \ .
\end{split}
\ee 
Meanwhile, the spectrum of this SCFT will contain the following chiral operators as a subset:
\begin{align}
\label{eq:ColBrAnDk}
\begin{split}
	&\left\{ {\rm Tr}\phi_{Sp,\ell}^{2k} \ ; \ 1 \leq \ell \leq m \ , \ N+1 \leq k \leq 2N\ell-N \right\} \  \cup \\
	&\left\{ {\rm Tr}\phi_{SO,\ell}^{2k} \ ; \ 1 \leq \ell \leq m-1 \ , \ N+1 \leq k \leq 2N \ell \right\} \ \cup \\
	&\left\{ {\rm Pf}\phi_{SO,\ell} \ ; \ 2 \leq \ell \leq m-1   \right\} \ \cup \\
	&\left\{ M_{j=2k-1} \ ; \ N+1 \leq k \leq 2m N   \right\} \cup  \\
	&\left\{M_{j=2mN} \ ; \ {\rm If } \ m>1   \right\} \ ,
\end{split}
\end{align}
with their corresponding dimensions being
\begin{align}
\begin{split}
	[{\rm Tr}\phi_{Sp,\ell}^{2k}] &= [{\rm Tr}\phi_{SO,\ell}^{2k}] = \frac{2k}{2N+1} \ , \\
	[{\rm Pf}\phi_{SO,\ell}]&=\frac{2Nl+1}{2N+1} \ , \\
	[M_j]&=\frac{j+1}{2N+1} \ .
\end{split}
\end{align}
Here $\phi_{Sp,\ell}$ is the adjoint chiral multiplet associated to the $\ell$-th symplectic gauge node. Similarly, $\phi_{SO,\ell}$ is the adjoint chiral multiplet associated to the $\ell$-th orthogonal gauge node and $M_j$ are gauge singlets forming the bottom most component of the spin-$j$ representation with respect to the $SU(2)$ embedding specified by $\rho$.

Based on the trend so far, it wouldn't be unreasonable to expect that the CFT at this fixed point has enhanced supersymmetry with \eqref{eq:ColBrAnDk} giving the spectrum of Coulomb branch operators. This is definitely true when $m=1$. In that case the above data matches that for the $(A_1, D_{2N-1})$ AD theory and the $\CN=1$ Lagrangian theory being described here coincides with that described in \cite{Agarwal:2016pjo}.   

As a more non-trivial example, consider the case when $N=1$ and $m=2$. The quiver in \eqref{eq:AnDkquiver}, then becomes
\begin{align}
\begin{split}
\boxed{SO(2)}-Sp(1)-SO(6)-Sp(3)-\boxed{SO(10)}  
\end{split}
\end{align}
Deforming this quiver in the manner described here, one can easily verify that $\eqref{eq:ColBrAnDk}$ will give us a set of $7$ chiral operators of dimensions  
\be
\left\{\frac{4}{3}, \frac{4}{3}, \frac{4}{3},\frac{5}{3}, 2,2,\frac{8}{3}     \right\} \ .
\ee 
This is identical to the list of Coulomb branch operators of the $(A_3,D_5)$ theory given in \eqref{eq:A3D5ColBr}. In fact, we checked that the operators listed in \eqref{eq:ColBrAnDk} are in one-to-one correspondence with the set of Coulomb branch operators of the $(A_{2m-1}, D_{2mN+1})$ AD theory. The central charges also agree with that of the above AD theories. We therefore conjecture that $\CN=1$ preserving principal nilpotent deformations of the quiver in \eqref{eq:AnDkquiver} trigger an RG flow that brings the theory to the fixed point described by the $(A_{2m-1}, D_{2mN+1})$ AD theory. 
 
 

\subsection{Lagrangian for $(A_{2m}, D_{m(N-2) + \frac{N}{2}})$ theory}

Let us now consider the quiver gauge theory where the left-most node carries $SO(N)$ gauge symmetry and containing a total of $2m$ gauge nodes, with $m$ of them carrying orthogonal gauge symmetry and $m$ of them carrying symplectic gauge symmetry. It can be represented by the following quiver diagram:
\begin{align}
\begin{split}
  SO(N)-Sp(N-2)-SO(3N-4)-Sp(2N-4)- ~~~~~~~~~~~~~~~~~~~~\\
 ~~~~~~~~~~~~~~~~ \ldots-Sp(m(N-2))-\boxed{SO(2m(N-2)+N)}  
\end{split}
\end{align}
The flavor symmetry of the quiver is then given by $SO(2m(N-2)+N)$. 

Now we subject these quivers to $\CN=1$ preserving nilpotent deformations, as was done in the previous sections. We find that the interacting theory at the IR fixed point always has rational central charges when the vev of the gauge singlet field $M$ corresponds to the principal nilpotent orbit of $SO(2m(N-2)+N)$ and $N$ is an even number greater than or equal to 4. The central charges for the IR theory are given by the following functions of $m$ and $N$:
\be
\begin{split} \label{bbb}
	a_{IR}&=\frac{m \left(4+16 m^2 (N-2)^2-13 N+8 N^2+24 m \left(2-3 N+N^2\right)\right)}{48 (N-1)} \ , \\
	c_{IR}&=\frac{m \left(4 m^2 (N-2)^2+N (-3+2 N)+6 m \left(2-3 N+N^2\right)\right)}{12 (N-1)} \ .
\end{split}
\ee 
At this fixed point, the spectrum will necessarily contain the following chiral operators:
\be
\begin{split}
	&\left\{ {\rm Tr}\phi_{Sp,\ell}^{2k} \ ; \ 1 \leq \ell \leq m \ , \ \frac{N}{2} \leq k \leq (N-2)\ell \right\} \  \cup \\
	&\left\{ {\rm Tr}\phi_{SO,\ell}^{2k} \ ; \ 1 \leq \ell \leq m \ , \ \frac{N}{2} \leq k \leq (N-2)\ell-\frac{N}{2}+1 \right\} \ \cup \\
	&\Big\{ {\rm Pf}\phi_{SO,\ell} \ ; \ 2 \leq \ell \leq m   \Big\} \ \cup \\
	&\left\{M_{j=(N-2)m+N/2-1} \ , \ M_{j=2k-1} \ ; \ \frac{N}{2} \leq k \leq \frac{(N-2)(2m+1) }{2}  \right \} \ , \\
\end{split}
\ee 
The respective IR dimensions of these operators are given by
\be
\label{eq:COAD}
\begin{split}
	[{\rm Tr}\phi_{Sp,\ell}^{2k}] &= [{\rm Tr}\phi_{SO,\ell}^{2k}] = \frac{2k}{N-1} \ , \\
	[{\rm Pf}\phi_{SO,\ell}]&=\frac{(2N-4)\ell-N+4}{2N-2} \ , \\
	[M_j]&=\frac{j+1}{N-1} \ .
\end{split}
\ee 
The data given above satisfies the relation given in \eqref{eq:STformula}, thereby providing a non-trivial consistency check for our conjecture that the $\mathcal{N} = 1$ principal nilpotent deformation of this quiver theory leads to an $\mathcal{N} = 2$ SCFT in the IR. 
In fact we checked that in all cases that we considered the above, the list of operators given in \eqref{eq:COAD} is in one-to-one correspondence with the spectrum of Coulomb branch operators of the $(A_{2m}, D_{m(N-2)+\frac{N}{2}})$ AD theory. The central charges \eqref{bbb} also agree with that of these AD theories. We therefore conjecture that the IR fixed point of our $\mathcal{N} = 1$ deformed quiver theory is described by the $(A_{2m}, D_{m(N-2)+\frac{N}{2}})$ AD theory.


An interesting observation 
is that when $N=4$, i.e. the starting node of the quiver has $SO(4)$ gauge symmetry, we have
\be
\label{eq:a=c}
a_{IR} = c_{IR} = \frac{1}{9} m \left(4 m^2+9 m+5\right) \ .
\ee 
Had we not been able to identify the IR theory, we might have wondered if the corresponding fixed points are actually some $\CN=3$ or $\CN=4$ theories, given that the two central charges become equal \cite{Aharony:2015oyb}. 
However, it is easy to exclude these possibilities. First of all, in all known 4d $\CN=4$ theories the central charges are such that $4a$ is integral, while this is not the case for the values listed in \eqref{eq:a=c} for generic $m$. Even if the central charges in \eqref{eq:a=c} do become integral when $m$ is $9k$ or $9k\pm1$, their putative Coulomb branch will contain operators with non-integral dimension: this is in contradiction with the fact that all the Coulomb branch operators in $\CN=4$ theories have integer dimension. 
Also, in all the known $\mathcal{N}=3$ theories as constructed in \cite{Garcia-Etxebarria:2015wns}, the Coulomb branch operators have integer dimensions \cite{Aharony:2016kai}. Therefore these couldn't have been $\mathcal{N}=3$ fixed points either.

\subsection{Lagrangian for $D_{m(2N+2)}^{m(2N+2)}[m]$ theory}

Finally, let us consider the quiver gauge theory starting with $Sp(N)$ gauge symmetry and containing a total of $2m-1$ gauge nodes, of which $m-1$ nodes carry an orthogonal gauge group and $m$ nodes carry a symplectic gauge group. It can be represented as the following quiver diagram: 
\begin{align}
\begin{split}
  Sp(N)-SO(4N+4)-Sp(3N+2)-SO(8N+8)- ~~~~~~~~~~~~~~~~~~\\
~~~~~~~~~~~~~~   \ldots-Sp((m-1)(2N+2)+N)-\boxed{SO((4N+4)m))}  
\end{split}
\end{align}
The flavor symmetry of this theory is given by $SO((4N+4)m)$. 

We now consider the $\CN=1$ deformation of this theory obtained by coupling a chiral superfield $M$ transforming in the adjoint representation of the flavor symmetry. This is done by adding the usual superpotential term given by
\be
\delta W = {\rm Tr}M \mu \ ,
\ee 
with $\mu$ being the moment map operator for the $SO((4N+4)m)$ flavor symmetry, and turning on the nilpotent vev specified by $\rho: SU(2)\hookrightarrow SO((4N+4)m)$. When $\rho$ corresponds to the principal embedding, we find that after removing all the operators and fields that decouple along the RG flow, the interacting fixed point has rational central charges which can be written as the following functions of $N$ and $m$: 
\begin{align}
\label{eq:centSPSO}
\begin{split}
	a_{IR} &= \frac{m \left(32 (m-1)^2 (1+N)^2+N (19+24 N)+8 (m-1) \left(5+13 N+8 N^2\right)\right)}{72+48 N} \ ,\\
	c_{IR} &= \frac{m \left(-2-5 N-2 N^2-6 m (1+N)+8 m^2 (1+N)^2\right)}{ 18+12 N} \ .
\end{split}
\end{align}
Given our past experience that whenever an $\CN=1$ nilpotent deformation consistently gave rational central charges, the RG flow always brought the theory to a fixed point with enhanced supersymmetry of Argyres-Douglas type, we believe that this must be true also in the present cases. 

Another reason to believe this is the fact that when $m=1$, the above central charges become
\begin{align}
\begin{split}
	a_{IR} &= \frac{N(19+24 N)}{72 + 48 N} \ ,\\
	c_{IR} &= \frac{N(5 + 6 N)}{18 + 12N} \ .
\end{split}
\end{align}
These are the central charges of the $(A_1, A_{2N} )$ theory, with the $m=1$ case having been already studied in \cite{Maruyoshi:2016aim}. We therefore believe that the above fixed points must correspond to some generalization of $(A_1,A_{2N})$ theories. After removing all the operators and fields that hit the unitarity bound, we find that the spectrum of chiral operators contains the following operators as a subset: 
\begin{align}
\begin{split}
	&\left\{ {\rm Tr}\phi_{Sp,\ell}^{2k} \ ; \ 2 \leq \ell \leq m \ , \ N+2 \leq k \leq (2N+2)(\ell-1) +N \right\} \  \cup \\
	&\left\{ {\rm Tr}\phi_{SO,\ell}^{2k} \ ; \ 1 \leq \ell \leq m-1 \ , \ N+2 \leq k \leq (2N+2)\ell-1 \right\} \ \cup \\
	&\left\{ {\rm Pf}\phi_{SO,\ell} \ ; \ 2 \leq \ell \leq m-1   \right\} \ \cup \\
	&\left\{M_{j=2k+1} \ ; \ N+1 \leq k \leq (2N +2)m-2   \right\} \cup \\
	&\left\{M_{j=(2N+2)m-1} \ ; \ {\rm If } \ m>1   \right\} \ ,
\end{split}
\end{align}
where we follow the same conventions for $\phi_{Sp,\ell} \ , \phi_{SO,\ell}$ and $M_j$ as in the previous subsection.

Once again, based on our experience, we expect that the above operators will be in one to one correspondence with the Coulomb branch operators of the $\CN=2$ theory that our fixed point corresponds to. Their respective dimensions are given by
\begin{align}
\begin{split}
[{\rm Tr}\phi_{Sp,\ell}^{2k}] &= [{\rm Tr}\phi_{SO,\ell}^{2k}] = \frac{2k}{2N+3} \ , \\
[{\rm Pf}\phi_{SO,\ell}] &=\frac{(2N+2)\ell}{2N+3} \ , \\
[M_j] &= \frac{j+1}{2N+3} \ . \label{33}
\end{split}
\end{align}
If this point has enhanced $\CN=2$ supersymmetry with the Coulomb branch operators given as above, it has to satisfy the relation given in \eqref{eq:STformula}.
We find that this is indeed true at the above fixed point. 

In fact in all the cases we checked, the list of operators given in \eqref{33} is in one-to-one correspondence with the spectrum of Coulomb branch operators of the $D_{m(2N+2)}^{m(2N+2)}[m]$ AD theory introduced in \cite{Wang:2015mra}. The central charges \eqref{eq:centSPSO} also agree with that of these AD theories. We therefore conjecture that the IR fixed point of our $\mathcal{N} = 1$ deformed quiver theory is described by the $D_{m(2N+2)}^{m(2N+2)}[m]$ AD theory.

\section{Conclusion} \label{sec:conclusion}

As a natural continuation of the works \cite{Maruyoshi:2016tqk, Maruyoshi:2016aim, Agarwal:2016pjo}, in this paper we studied particular $\mathcal{N} = 1$ preserving deformations of a class of $\mathcal{N} = 2$ superconformal \textit{quiver} gauge theories labelled by the $SU(2)$ embedding $\rho$ into the (subgroup) of the flavor symmetry.  
More precisely, the quiver theories we considered are built out of $SU$ nodes (quivers in \eqref{eq:SUquiver}) or alternating $SO - Sp$ nodes (quivers in \eqref{eq:2ndSPSO}, \eqref{eq:SOSP1}, \eqref{eq:SPSO1}). 
When $\rho$ corresponds to a non-principal embedding, we found that these deformations seem to lead in general to $\mathcal{N} = 1$ SCFTs in the IR; although it may be possible that at least some of these $\mathcal{N} = 1$ SCFTs are already known in the literature, stating a precise relation is quite hard at the moment.
On the other hand, when $\rho$ is chosen to be the principal embedding, the 
$\mathcal{N} = 1$ deformation triggers an RG flow to IR superconformal theory with enhanced $\mathcal{N} = 2$ supersymmetry of Argyres-Douglas type.
This led us to find UV Lagrangian descriptions for the AD theories of type $(A_{k-1}, A_{mk-1})$, $(I_{m,mk}, S)$, $(A_{2m-1}, D_{2Nm+1})$, $(A_{2m}, D_{m(N-2) + \frac{N}{2}})$ (for $N$ even) and $D_{m(2N+2)}^{m(2N+2)}[m]$. 
Interestingly, as already observed in \cite{Maruyoshi:2016tqk, Maruyoshi:2016aim, Agarwal:2016pjo} $\mathcal{N} = 2$ SUSY enhancement seems to happen only when the flavor group of the Lagrangian UV theory is of ADE type; however at the moment we don't have a clear understanding of why this should be the case.

The $\CN=2$ quiver gauge theories we ``$\CN=1$ deformed'' to obtain AD theories have one feature in common. Their associated chiral algebras (in the sense of \cite{Beem:2013sza}) satisfy the Sugawara relation:
\begin{align}
 c_{2d} = \frac{k_{2d} \textrm{dim} G_F}{k_{2d} + h^\vee}  \,.
\end{align}
Here $c_{2d} = -12 c_{4d}$ and $k_{2d} = -\half k_{4d}$ with $k_{4d}$ being the flavor central charge associated to the flavor symmetry $G_F$. For the $SU(n)$ quiver theories that have $U(1)$ factors, we also need to sum up these contributions ($c_{2d}=1$ for each $U(1)$'s). This is consistent with the conjecture made in \cite{Agarwal:2016pjo}. We also find that the flavor central charge is strictly above the bound $k_{4d} \ge k_{\textrm{bound}}$ of \cite{Beem:2013sza,Lemos:2015orc} (except when there is only one gauge factor in the quiver). This is also consistent with the conjecture that only the principal nilpotent vev will trigger a SUSY enhancing flow unless the bound is saturated. 

There are a number of open questions and interesting directions yet to be explored.
An immediate application is to use our Lagrangian descriptions to compute the superconformal index or other supersymmetric partition functions for the Argyres-Douglas theories we found in this paper. The Schur index for the $(A_{k-1}, A_{mk-1})$ and $(I_{m, mk}, S)$ theory has been recently obtained in \cite{Buican:2017uka}. It should be possible to compare directly with this result and also give a further prediction of the full index. Moreover, we can give a prediction for the indices of $(A_{m-1}, D_{2Nm+1})$, $(A_{2m}, D_{m(N-2) + \frac{N}{2}})$ and $D_{m(2N+2)}^{m(2N+2)}[m]$ theories. 
The IR computation of the Schur index along the line of \cite{Cordova:2015nma} will provide an independent check. 

It would also be important to look for more examples, \textit{i.e.} to study deformations of a larger class of Lagrangian theories.  An exciting possibility would be to find additional IR non-Lagrangian SCFTs with $a=c$ like the ones we obtained in Section \ref{sec:SPSOquiver}, but with enhanced $\mathcal{N} = 3$ or $\mathcal{N} = 4$ supersymmetry, although it is not yet clear to us if this is possible by using this kind of $\mathcal{N} = 1$ deformations.
In any case, having more examples will also hopefully lead us to understand
which Lagrangian theories will have a chance to flow to an AD theory, or to some other interesting non-Lagrangian SCFTs; this is related to the problem of better understanding the mechanism at work in the $\mathcal{N} = 1$ deformation introduced in \cite{Maruyoshi:2016tqk, Maruyoshi:2016aim, Agarwal:2016pjo}, which is still rather mysterious. It will also be interesting to explore these RG flows using the methods developed in \cite{Gukov:2015qea, Gukov:2016tnp}.

Another direction would be to understand how and if $S$-duality of $\mathcal{N}=2$ quiver theories can be related to $S$-duality for the AD theories obtained after the $\mathcal{N}=1$ deformation. We are currently investigating this point.

Finally, it would be very interesting to study the reduction of our four-dimensional quiver Lagrangian theories to three dimensions along the lines of \cite{Benvenuti:2017kud,Benvenuti:2017lle}, which would have an implication on the 3d mirror symmetry. 
It should be possible to push the dimensional reduction further to two dimensions along the lines of \cite{Putrov:2015jpa, Cecotti:2015lab, Honda:2015yha, Gadde:2015wta, Amariti:2017cyd, Gukov:2017zao}.

We hope to be able to address some of these points in the near future.

\begin{acknowledgments}
We thank Sergio Cecotti, Dongmin Gang and Kazunobu Maruyoshi for many helpful discussions.
The work of PA is supported in part by Samsung Science and Technology Foundation under Project Number SSTF-BA1402-08, in part by National Research Foundation of Korea grant number 2015R1A2A2A01003124 and in part by the Korea Research Fellowship Program through the National Research Foundation of Korea funded by the Ministry of Science, ICT and Future Planning, grant number 2016H1D3A1938054.
AS would like to thank the organizers of the workshop `Geometric Correspondences of Gauge Theories' at SISSA-ICTP for hospitality during the completion of this work.
JS would like to thank the organizers of the workshop `Geometry of String and Gauge Theories' at CERN, and also CERN-Korea Theory Collaboration funded by National Research Foundation (Korea) for the hospitality and support. 

\end{acknowledgments}

\bibliographystyle{jhep}
\bibliography{ADN1}

\providecommand{\href}[2]{#2}\begingroup\raggedright\begin{thebibliography}{10}

\bibitem{Argyres:1995jj}
P.~C. Argyres and M.~R. Douglas, \emph{{New phenomena in SU(3) supersymmetric
  gauge theory}},
  \href{https://doi.org/10.1016/0550-3213(95)00281-V}{\emph{Nucl. Phys.}
  {\bfseries B448} (1995) 93--126},
  [\href{https://arxiv.org/abs/hep-th/9505062}{{\ttfamily hep-th/9505062}}].

\bibitem{Argyres:1995xn}
P.~C. Argyres, M.~R. Plesser, N.~Seiberg and E.~Witten, \emph{{New N=2
  superconformal field theories in four-dimensions}},
  \href{https://doi.org/10.1016/0550-3213(95)00671-0}{\emph{Nucl. Phys.}
  {\bfseries B461} (1996) 71--84},
  [\href{https://arxiv.org/abs/hep-th/9511154}{{\ttfamily hep-th/9511154}}].

\bibitem{Eguchi:1996ds}
T.~Eguchi and K.~Hori, \emph{{N=2 superconformal field theories in
  four-dimensions and A-D-E classification}},  in \emph{{The mathematical
  beauty of physics: A memorial volume for Claude Itzykson. Proceedings,
  Conference, Saclay, France, June 5-7, 1996}}, pp.~67--82, 1996,
  \href{https://arxiv.org/abs/hep-th/9607125}{{\ttfamily hep-th/9607125}}.

\bibitem{Eguchi:1996vu}
T.~Eguchi, K.~Hori, K.~Ito and S.-K. Yang, \emph{{Study of N=2 superconformal
  field theories in four-dimensions}},
  \href{https://doi.org/10.1016/0550-3213(96)00188-5}{\emph{Nucl. Phys.}
  {\bfseries B471} (1996) 430--444},
  [\href{https://arxiv.org/abs/hep-th/9603002}{{\ttfamily hep-th/9603002}}].

\bibitem{Cecotti:2010fi}
S.~Cecotti, A.~Neitzke and C.~Vafa, \emph{{R-Twisting and 4d/2d
  Correspondences}},  \href{https://arxiv.org/abs/1006.3435}{{\ttfamily
  1006.3435}}.

\bibitem{Xie:2012hs}
D.~Xie, \emph{{General Argyres-Douglas Theory}},
  \href{https://doi.org/10.1007/JHEP01(2013)100}{\emph{JHEP} {\bfseries 1301}
  (2013) 100}, [\href{https://arxiv.org/abs/1204.2270}{{\ttfamily 1204.2270}}].

\bibitem{Cecotti:2012jx}
S.~Cecotti and M.~Del~Zotto, \emph{{Infinitely many N=2 SCFT with ADE flavor
  symmetry}}, \href{https://doi.org/10.1007/JHEP01(2013)191}{\emph{JHEP}
  {\bfseries 01} (2013) 191},
  [\href{https://arxiv.org/abs/1210.2886}{{\ttfamily 1210.2886}}].

\bibitem{Cecotti:2013lda}
S.~Cecotti, M.~Del~Zotto and S.~Giacomelli, \emph{{More on the N=2
  superconformal systems of type $D_p(G)$}},
  \href{https://doi.org/10.1007/JHEP04(2013)153}{\emph{JHEP} {\bfseries 04}
  (2013) 153}, [\href{https://arxiv.org/abs/1303.3149}{{\ttfamily 1303.3149}}].

\bibitem{Wang:2015mra}
Y.~Wang and D.~Xie, \emph{{Classification of Argyres-Douglas theories from M5
  branes}}, \href{https://doi.org/10.1103/PhysRevD.94.065012}{\emph{Phys. Rev.}
  {\bfseries D94} (2016) 065012},
  [\href{https://arxiv.org/abs/1509.00847}{{\ttfamily 1509.00847}}].

\bibitem{Xie:2016uqq}
D.~Xie and S.-T. Yau, \emph{{New N = 2 dualities}},
  \href{https://arxiv.org/abs/1602.03529}{{\ttfamily 1602.03529}}.

\bibitem{Xie:2017vaf}
D.~Xie and S.-T. Yau, \emph{{Argyres-Douglas matter and N=2 dualities}},
  \href{https://arxiv.org/abs/1701.01123}{{\ttfamily 1701.01123}}.

\bibitem{Maruyoshi:2016tqk}
K.~Maruyoshi and J.~Song, \emph{{Enhancement of Supersymmetry via
  Renormalization Group Flow and the Superconformal Index}},
  \href{https://doi.org/10.1103/PhysRevLett.118.151602}{\emph{Phys. Rev. Lett.}
  {\bfseries 118} (2017) 151602},
  [\href{https://arxiv.org/abs/1606.05632}{{\ttfamily 1606.05632}}].

\bibitem{Maruyoshi:2016aim}
K.~Maruyoshi and J.~Song, \emph{{$ \mathcal{N}=1 $ deformations and RG flows of
  $ \mathcal{N}=2 $ SCFTs}},
  \href{https://doi.org/10.1007/JHEP02(2017)075}{\emph{JHEP} {\bfseries 02}
  (2017) 075}, [\href{https://arxiv.org/abs/1607.04281}{{\ttfamily
  1607.04281}}].

\bibitem{Agarwal:2016pjo}
P.~Agarwal, K.~Maruyoshi and J.~Song, \emph{{$ \mathcal{N} $ =1 Deformations
  and RG flows of $ \mathcal{N} $ =2 SCFTs, part II: non-principal
  deformations}}, \href{https://doi.org/10.1007/JHEP12(2016)103,
  10.1007/JHEP04(2017)113}{\emph{JHEP} {\bfseries 12} (2016) 103},
  [\href{https://arxiv.org/abs/1610.05311}{{\ttfamily 1610.05311}}].

\bibitem{Gadde:2013fma}
A.~Gadde, K.~Maruyoshi, Y.~Tachikawa and W.~Yan, \emph{{New N=1 Dualities}},
  \href{https://doi.org/10.1007/JHEP06(2013)056}{\emph{JHEP} {\bfseries 1306}
  (2013) 056}, [\href{https://arxiv.org/abs/1303.0836}{{\ttfamily 1303.0836}}].

\bibitem{Agarwal:2013uga}
P.~Agarwal and J.~Song, \emph{{New N=1 Dualities from M5-branes and
  Outer-automorphism Twists}},
  \href{https://doi.org/10.1007/JHEP03(2014)133}{\emph{JHEP} {\bfseries 03}
  (2014) 133}, [\href{https://arxiv.org/abs/1311.2945}{{\ttfamily 1311.2945}}].

\bibitem{Agarwal:2014rua}
P.~Agarwal, I.~Bah, K.~Maruyoshi and J.~Song, \emph{{Quiver tails and $
  \mathcal{N}=1 $ SCFTs from M5-branes}},
  \href{https://doi.org/10.1007/JHEP03(2015)049}{\emph{JHEP} {\bfseries 03}
  (2015) 049}, [\href{https://arxiv.org/abs/1409.1908}{{\ttfamily 1409.1908}}].

\bibitem{Giacomelli:2014rna}
S.~Giacomelli, \emph{{Four dimensional superconformal theories from M5
  branes}}, \href{https://doi.org/10.1007/JHEP01(2015)044}{\emph{JHEP}
  {\bfseries 01} (2015) 044},
  [\href{https://arxiv.org/abs/1409.3077}{{\ttfamily 1409.3077}}].

\bibitem{Agarwal:2015vla}
P.~Agarwal, K.~Intriligator and J.~Song, \emph{{Infinitely many $ \mathcal{N}=1
  $ dualities from m + 1 − m = 1}},
  \href{https://doi.org/10.1007/JHEP10(2015)035}{\emph{JHEP} {\bfseries 10}
  (2015) 035}, [\href{https://arxiv.org/abs/1505.00255}{{\ttfamily
  1505.00255}}].

\bibitem{Fazzi:2016eec}
M.~Fazzi and S.~Giacomelli, \emph{{$\mathcal{N} = 1$ superconformal theories
  with $D_N$ blocks}},
  \href{https://doi.org/10.1103/PhysRevD.95.085010}{\emph{Phys. Rev.}
  {\bfseries D95} (2017) 085010},
  [\href{https://arxiv.org/abs/1609.08156}{{\ttfamily 1609.08156}}].

\bibitem{Nardoni:2016ffl}
E.~Nardoni, \emph{{4d SCFTs from negative-degree line bundles}},
  \href{https://arxiv.org/abs/1611.01229}{{\ttfamily 1611.01229}}.

\bibitem{Intriligator:2003jj}
K.~A. Intriligator and B.~Wecht, \emph{{The Exact superconformal R symmetry
  maximizes a}},
  \href{https://doi.org/10.1016/S0550-3213(03)00459-0}{\emph{Nucl. Phys.}
  {\bfseries B667} (2003) 183--200},
  [\href{https://arxiv.org/abs/hep-th/0304128}{{\ttfamily hep-th/0304128}}].

\bibitem{Kutasov:2003iy}
D.~Kutasov, A.~Parnachev and D.~A. Sahakyan, \emph{{Central charges and U(1)(R)
  symmetries in N=1 superYang-Mills}},
  \href{https://doi.org/10.1088/1126-6708/2003/11/013}{\emph{JHEP} {\bfseries
  11} (2003) 013}, [\href{https://arxiv.org/abs/hep-th/0308071}{{\ttfamily
  hep-th/0308071}}].

\bibitem{Benvenuti:2017kud}
S.~Benvenuti and S.~Giacomelli, \emph{{Abelianization and Sequential
  Confinement in $2+1$ dimensions}},
  \href{https://arxiv.org/abs/1706.04949}{{\ttfamily 1706.04949}}.

\bibitem{Benvenuti:2017lle}
S.~Benvenuti and S.~Giacomelli, \emph{{Compactification of dualities with
  decoupled operators and $3d$ mirror symmetry}},
  \href{https://arxiv.org/abs/1706.02225}{{\ttfamily 1706.02225}}.

\bibitem{Buican:2015ina}
M.~Buican and T.~Nishinaka, \emph{{On the superconformal index of
  Argyres--Douglas theories}},
  \href{https://doi.org/10.1088/1751-8113/49/1/015401}{\emph{J. Phys.}
  {\bfseries A49} (2016) 015401},
  [\href{https://arxiv.org/abs/1505.05884}{{\ttfamily 1505.05884}}].

\bibitem{Buican:2015tda}
M.~Buican and T.~Nishinaka, \emph{{Argyres-Douglas Theories, the Macdonald
  Index, and an RG Inequality}},
  \href{https://doi.org/10.1007/JHEP02(2016)159}{\emph{JHEP} {\bfseries 02}
  (2016) 159}, [\href{https://arxiv.org/abs/1509.05402}{{\ttfamily
  1509.05402}}].

\bibitem{Song:2015wta}
J.~Song, \emph{{Superconformal indices of generalized Argyres-Douglas theories
  from 2d TQFT}}, \href{https://doi.org/10.1007/JHEP02(2016)045}{\emph{JHEP}
  {\bfseries 02} (2016) 045},
  [\href{https://arxiv.org/abs/1509.06730}{{\ttfamily 1509.06730}}].

\bibitem{Buican:2017uka}
M.~Buican and T.~Nishinaka, \emph{{On Irregular Singularity Wave Functions and
  Superconformal Indices}},  \href{https://arxiv.org/abs/1705.07173}{{\ttfamily
  1705.07173}}.

\bibitem{Song:2017oew}
J.~Song, D.~Xie and W.~Yan, \emph{{Vertex operator algebras of Argyres-Douglas
  theories from M5-branes}},
  \href{https://arxiv.org/abs/1706.01607}{{\ttfamily 1706.01607}}.

\bibitem{DelZotto:2014kka}
M.~Del~Zotto and A.~Hanany, \emph{{Complete Graphs, Hilbert Series, and the
  Higgs branch of the 4d N=2 $(A_n,A_m)$ SCFT's}},
  \href{https://arxiv.org/abs/1403.6523}{{\ttfamily 1403.6523}}.

\bibitem{Cecotti:2015lab}
S.~Cecotti, J.~Song, C.~Vafa and W.~Yan, \emph{{Superconformal Index, BPS
  Monodromy and Chiral Algebras}},
  \href{https://arxiv.org/abs/1511.01516}{{\ttfamily 1511.01516}}.

\bibitem{Cordova:2015nma}
C.~Cordova and S.-H. Shao, \emph{{Schur Indices, BPS Particles, and
  Argyres-Douglas Theories}},
  \href{https://doi.org/10.1007/JHEP01(2016)040}{\emph{JHEP} {\bfseries 01}
  (2016) 040}, [\href{https://arxiv.org/abs/1506.00265}{{\ttfamily
  1506.00265}}].

\bibitem{Cordova:2016uwk}
C.~Cordova, D.~Gaiotto and S.-H. Shao, \emph{{Infrared Computations of Defect
  Schur Indices}}, \href{https://doi.org/10.1007/JHEP11(2016)106}{\emph{JHEP}
  {\bfseries 11} (2016) 106},
  [\href{https://arxiv.org/abs/1606.08429}{{\ttfamily 1606.08429}}].

\bibitem{Cordova:2017ohl}
C.~Cordova, D.~Gaiotto and S.-H. Shao, \emph{{Surface Defect Indices and 2d-4d
  BPS States}},  \href{https://arxiv.org/abs/1703.02525}{{\ttfamily
  1703.02525}}.

\bibitem{Beem:2013sza}
C.~Beem, M.~Lemos, P.~Liendo, W.~Peelaers, L.~Rastelli and B.~C. van Rees,
  \emph{{Infinite Chiral Symmetry in Four Dimensions}},
  \href{https://doi.org/10.1007/s00220-014-2272-x}{\emph{Commun. Math. Phys.}
  {\bfseries 336} (2015) 1359--1433},
  [\href{https://arxiv.org/abs/1312.5344}{{\ttfamily 1312.5344}}].

\bibitem{Xie:2016evu}
D.~Xie, W.~Yan and S.-T. Yau, \emph{{Chiral algebra of Argyres-Douglas theory
  from M5 brane}},  \href{https://arxiv.org/abs/1604.02155}{{\ttfamily
  1604.02155}}.

\bibitem{Song:2016yfd}
J.~Song, \emph{{Macdonald Index and Chiral Algebra}},
  \href{https://arxiv.org/abs/1612.08956}{{\ttfamily 1612.08956}}.

\bibitem{Creutzig:2017qyf}
T.~Creutzig, \emph{{W-algebras for Argyres-Douglas theories}},
  \href{https://arxiv.org/abs/1701.05926}{{\ttfamily 1701.05926}}.

\bibitem{Cordova:2017mhb}
C.~Cordova, D.~Gaiotto and S.-H. Shao, \emph{{Surface Defects and Chiral
  Algebras}}, \href{https://doi.org/10.1007/JHEP05(2017)140}{\emph{JHEP}
  {\bfseries 05} (2017) 140},
  [\href{https://arxiv.org/abs/1704.01955}{{\ttfamily 1704.01955}}].

\bibitem{Fredrickson:2017yka}
L.~Fredrickson, D.~Pei, W.~Yan and K.~Ye, \emph{{Argyres-Douglas Theories,
  Chiral Algebras and Wild Hitchin Characters}},
  \href{https://arxiv.org/abs/1701.08782}{{\ttfamily 1701.08782}}.

\bibitem{Gukov:2017zao}
S.~Gukov, \emph{{Trisecting non-Lagrangian theories}},
  \href{https://arxiv.org/abs/1707.01515}{{\ttfamily 1707.01515}}.

\bibitem{Giacomelli:2017ckh}
S.~Giacomelli, \emph{{RG flows with supersymmetry enhancement and geometric
  engineering}},  \href{https://arxiv.org/abs/1710.06469}{{\ttfamily
  1710.06469}}.

\bibitem{Benvenuti:2017bpg}
S.~Benvenuti and S.~Giacomelli, \emph{{Lagrangians for generalized
  Argyres-Douglas theories}},
  \href{https://arxiv.org/abs/1707.05113}{{\ttfamily 1707.05113}}.

\bibitem{Gaiotto:2009hg}
D.~Gaiotto, G.~W. Moore and A.~Neitzke, \emph{{Wall-crossing, Hitchin Systems,
  and the WKB Approximation}},
  \href{https://arxiv.org/abs/0907.3987}{{\ttfamily 0907.3987}}.

\bibitem{Gaiotto:2009we}
D.~Gaiotto, \emph{{N=2 dualities}},
  \href{https://doi.org/10.1007/JHEP08(2012)034}{\emph{JHEP} {\bfseries 1208}
  (2012) 034}, [\href{https://arxiv.org/abs/0904.2715}{{\ttfamily 0904.2715}}].

\bibitem{Shapere:1999xr}
A.~D. Shapere and C.~Vafa, \emph{{BPS structure of Argyres-Douglas
  superconformal theories}},
  \href{https://arxiv.org/abs/hep-th/9910182}{{\ttfamily hep-th/9910182}}.

\bibitem{Buican:2014hfa}
M.~Buican, S.~Giacomelli, T.~Nishinaka and C.~Papageorgakis,
  \emph{{Argyres-Douglas Theories and S-Duality}},
  \href{https://doi.org/10.1007/JHEP02(2015)185}{\emph{JHEP} {\bfseries 02}
  (2015) 185}, [\href{https://arxiv.org/abs/1411.6026}{{\ttfamily 1411.6026}}].

\bibitem{Buican:2017fiq}
M.~Buican, Z.~Laczko and T.~Nishinaka, \emph{{N=2 S-duality Revisited}},
  \href{https://arxiv.org/abs/1706.03797}{{\ttfamily 1706.03797}}.

\bibitem{Shapere:2008zf}
A.~D. Shapere and Y.~Tachikawa, \emph{{Central charges of N=2 superconformal
  field theories in four dimensions}},
  \href{https://doi.org/10.1088/1126-6708/2008/09/109}{\emph{JHEP} {\bfseries
  09} (2008) 109}, [\href{https://arxiv.org/abs/0804.1957}{{\ttfamily
  0804.1957}}].

\bibitem{Aharony:2016kai}
O.~Aharony and Y.~Tachikawa, \emph{{S-folds and 4d N=3 superconformal field
  theories}}, \href{https://doi.org/10.1007/JHEP06(2016)044}{\emph{JHEP}
  {\bfseries 06} (2016) 044},
  [\href{https://arxiv.org/abs/1602.08638}{{\ttfamily 1602.08638}}].

\bibitem{Argyres:2016yzz}
P.~C. Argyres and M.~Martone, \emph{{4d $ \mathcal{N} $ =2 theories with
  disconnected gauge groups}},
  \href{https://doi.org/10.1007/JHEP03(2017)145}{\emph{JHEP} {\bfseries 03}
  (2017) 145}, [\href{https://arxiv.org/abs/1611.08602}{{\ttfamily
  1611.08602}}].

\bibitem{Xie:2013jc}
D.~Xie and P.~Zhao, \emph{{Central charges and RG flow of strongly-coupled N=2
  theory}}, \href{https://doi.org/10.1007/JHEP03(2013)006}{\emph{JHEP}
  {\bfseries 03} (2013) 006},
  [\href{https://arxiv.org/abs/1301.0210}{{\ttfamily 1301.0210}}].

\bibitem{Morita:2011cs}
T.~Morita and V.~Niarchos, \emph{{F-theorem, duality and SUSY breaking in
  one-adjoint Chern-Simons-Matter theories}},
  \href{https://doi.org/10.1016/j.nuclphysb.2012.01.003}{\emph{Nucl. Phys.}
  {\bfseries B858} (2012) 84--116},
  [\href{https://arxiv.org/abs/1108.4963}{{\ttfamily 1108.4963}}].

\bibitem{Agarwal:2012wd}
P.~Agarwal, A.~Amariti and M.~Siani, \emph{{Refined Checks and Exact Dualities
  in Three Dimensions}},
  \href{https://doi.org/10.1007/JHEP10(2012)178}{\emph{JHEP} {\bfseries 10}
  (2012) 178}, [\href{https://arxiv.org/abs/1205.6798}{{\ttfamily 1205.6798}}].

\bibitem{Safdi:2012re}
B.~R. Safdi, I.~R. Klebanov and J.~Lee, \emph{{A Crack in the Conformal
  Window}}, \href{https://doi.org/10.1007/JHEP04(2013)165}{\emph{JHEP}
  {\bfseries 04} (2013) 165},
  [\href{https://arxiv.org/abs/1212.4502}{{\ttfamily 1212.4502}}].

\bibitem{Aharony:2015oyb}
O.~Aharony and M.~Evtikhiev, \emph{{On four dimensional N = 3 superconformal
  theories}}, \href{https://doi.org/10.1007/JHEP04(2016)040}{\emph{JHEP}
  {\bfseries 04} (2016) 040},
  [\href{https://arxiv.org/abs/1512.03524}{{\ttfamily 1512.03524}}].

\bibitem{Garcia-Etxebarria:2015wns}
I.~García-Etxebarria and D.~Regalado, \emph{{$ \mathcal{N}=3 $ four
  dimensional field theories}},
  \href{https://doi.org/10.1007/JHEP03(2016)083}{\emph{JHEP} {\bfseries 03}
  (2016) 083}, [\href{https://arxiv.org/abs/1512.06434}{{\ttfamily
  1512.06434}}].

\bibitem{Lemos:2015orc}
M.~Lemos and P.~Liendo, \emph{{$\mathcal{N}=2$ central charge bounds from $2d$
  chiral algebras}}, \href{https://doi.org/10.1007/JHEP04(2016)004}{\emph{JHEP}
  {\bfseries 04} (2016) 004},
  [\href{https://arxiv.org/abs/1511.07449}{{\ttfamily 1511.07449}}].

\bibitem{Gukov:2015qea}
S.~Gukov, \emph{{Counting RG flows}},
  \href{https://doi.org/10.1007/JHEP01(2016)020}{\emph{JHEP} {\bfseries 01}
  (2016) 020}, [\href{https://arxiv.org/abs/1503.01474}{{\ttfamily
  1503.01474}}].

\bibitem{Gukov:2016tnp}
S.~Gukov, \emph{{RG Flows and Bifurcations}},
  \href{https://doi.org/10.1016/j.nuclphysb.2017.03.025}{\emph{Nucl. Phys.}
  {\bfseries B919} (2017) 583--638},
  [\href{https://arxiv.org/abs/1608.06638}{{\ttfamily 1608.06638}}].

\bibitem{Putrov:2015jpa}
P.~Putrov, J.~Song and W.~Yan, \emph{{(0,4) dualities}},
  \href{https://doi.org/10.1007/JHEP03(2016)185}{\emph{JHEP} {\bfseries 03}
  (2016) 185}, [\href{https://arxiv.org/abs/1505.07110}{{\ttfamily
  1505.07110}}].

\bibitem{Honda:2015yha}
M.~Honda and Y.~Yoshida, \emph{{Supersymmetric index on $T^2 x S^2$ and
  elliptic genus}},  \href{https://arxiv.org/abs/1504.04355}{{\ttfamily
  1504.04355}}.

\bibitem{Gadde:2015wta}
A.~Gadde, S.~S. Razamat and B.~Willett, \emph{{On the reduction of 4d $
  \mathcal{N}=1 $ theories on $ {\mathbb{S}}^2 $}},
  \href{https://doi.org/10.1007/JHEP11(2015)163}{\emph{JHEP} {\bfseries 11}
  (2015) 163}, [\href{https://arxiv.org/abs/1506.08795}{{\ttfamily
  1506.08795}}].

\bibitem{Amariti:2017cyd}
A.~Amariti, L.~Cassia and S.~Penati, \emph{{Surveying 4d SCFTs twisted on
  Riemann surfaces}},
  \href{https://doi.org/10.1007/JHEP06(2017)056}{\emph{JHEP} {\bfseries 06}
  (2017) 056}, [\href{https://arxiv.org/abs/1703.08201}{{\ttfamily
  1703.08201}}].

\end{thebibliography}\endgroup

\end{document}